\documentclass[aps,prl,reprint,groupedaddress,amsmath,amssymb,superscriptaddress]{revtex4-2}
\usepackage{graphicx}
\usepackage{float}
\usepackage{multirow}
\usepackage{natbib,twoopt}
\usepackage[breaklinks=true]{hyperref}
\usepackage{xcolor}
\usepackage{amssymb}
\usepackage{stackengine,graphicx}
\usepackage{amsmath}
\usepackage{hyperref}
\usepackage{color}
\usepackage{stackengine}
\usepackage{subfigure}
\usepackage{verbatim}


\newcommand{\RNum}[1]{\uppercase\expandafter{\romannumeral #1\relax}}

\newcommand{\balancecolsandclearpage}{%
	\close@column@grid
	\cleardoublepage
	\twocolumngrid
}

\begin{document} 

\title{On anomalous optical beam shifts at near-normal incidence}
	
\author{Maxim Mazanov}
\affiliation{V. N. Karazin Kharkiv National University, Kharkiv, Ukraine}
\affiliation{School of Physics and Engineering, ITMO University, St. Petersburg, Russia}
\author{Oleh Yermakov}
\email{oe.yermakov@gmail.com}
\affiliation{V. N. Karazin Kharkiv National University, Kharkiv, Ukraine}
\affiliation{School of Physics and Engineering, ITMO University, St. Petersburg, Russia}
\author{Andrey Bogdanov}
\affiliation{School of Physics and Engineering, ITMO University, St. Petersburg, Russia}
\author{Andrei Lavrinenko}
\affiliation{Department of Photonics Engineering, Technical University of Denmark, Kgs. Lyngby, Denmark}
  
\begin{abstract}

We develop the theory of optical beam shifts (both Goos-H{\"a}nchen and Imbert-Fedorov) for the case of near-normal incidence, when the incident angle becomes comparable with the angular beam divergence. Such a situation naturally leads to strong enhancement of the shifts reported recently [ACS Photonics 6, 2530 (2019)]. Experimental results find complete and rigorous explanation in our generalized theory. In addition, the developed theory uncovers the unified origin of the anomalous beam shifts enhancement via the Berry phase singularity. We also propose a simple experimental scheme involving quarter-wave plate that allows to observe the giant transverse and longitudinal, spatial and angular beam shifts simultaneously. Our results can find applications in spin-orbit photonics, polarization optics, sensing applications, and quantum weak measurements.

\end{abstract}


\maketitle


The reflection and refraction of a plane electromagnetic wave at a dielectric interface are the basic physical processes inherent to all-optical systems and devices. They are rigorously described by the Snell's law and Fresnel equations. Nevertheless, in practice, we usually deal with optical beams of a finite width for which the plane wave approximation is oversimplified. This results in the deviation from geometrical optics that manifests itself as spatial shifts of the beam known as the Goos-H{\"a}nchen (GH) and Imbert-Fedorov (IF) shifts. These shifts can be explained in terms of weak material-mediated interaction of the beam spectrum ({\it orbit state}) and its polarization ({\it spin state})~\cite{dennis2012analogy,gotte2012generalized,toppel2013goos,bliokh2013goos}.

Today, the GH and IF shifts, including the photonic spin Hall effect (PSHE) as a particular example of IF shift, are well-studied in different material systems including atomic optics, optical sensors, graphene, metasurfaces, polarizers, uniaxial crystals, etc~\cite{menzel2008imbert,yin2013photonic,bliokh2016spin,Ling2017}. Usually, the spatial (angular) beam shifts are very small -- typically of the order of the light wavelength (the beam angular spectrum variance) -- that limits their application \cite{bliokh2013goos,dennis2012analogy,gotte2012generalized,toppel2013goos}. The shifts can be enhanced under several specific conditions including the near-Brewster incidence~\cite{gotte2014eigenpolarizations,gotte2013limits,aiello2009brewster,merano2009observing}, material resonances~\cite{soboleva2012giant,salasnich2012enhancement}, exceptional points~\cite{zhou2019controlling}, and output beam polarization post-selection~\cite{hosten2008observation}. In all these cases, the enhancement occurs at large angles of incidence (typically, $>10^\circ$). However, the anomalous enhancement of the PSHE in hyperbolic metamaterial slabs has been experimentally observed recently at a near-normal angle of incidence~\cite{takayama2018photonic,kim2019observation}. This is very counter-intuitive as at the strictly normal incidence the PSHE completely disappears. 
Therefore, at first sight, there are no rigorous arguments for the enhancement of PSHE at small angles of incidence. Moreover, the standard theory of optical beam shifts~\cite{bliokh2013goos} fails to explain the observed effect.

In this Letter, we develop a generalized theory of beam shifts addressing small incident angles. We show that anomalously large spatial and angular shifts of both types (GH and IF) could be simultaneously observed for the beam near-normally transmitted or reflected at an uniaxial slab. Usually, the longitudinal (GH) and transverse (IF) shifts are caused by different origins and considered independently. Here, we show that anomalous GH and IF shifts have the same nature caused by the Berry phase singularity
when the incident angle becomes comparable to the angular divergence of the beam. 
While the partial cases of near-normal IF shift enhancement have been observed~\cite{takayama2018photonic,kim2019observation,Zhu2021PSHE,Ling2021}, we predict another intriguing feature, namely the anomalous GH shift which is commonly known to exist under total internal reflection conditions~\cite{bliokh2013goos}. In the frame of the developed generalized theory, we analyze necessary and optimum conditions for the giant beam shifts of all types (GH and IF, linear and angular) and reveal that a conventional low-birefringent quarter-wave plate (QWP) can be used in such a remarkably simple experiment. 
Finally, we analyze the relevant beam parameters, polarization structure and intensity profiles of the shifted transmitted beam.

\begin{figure}[h!]
	\includegraphics [width=0.45\textwidth]{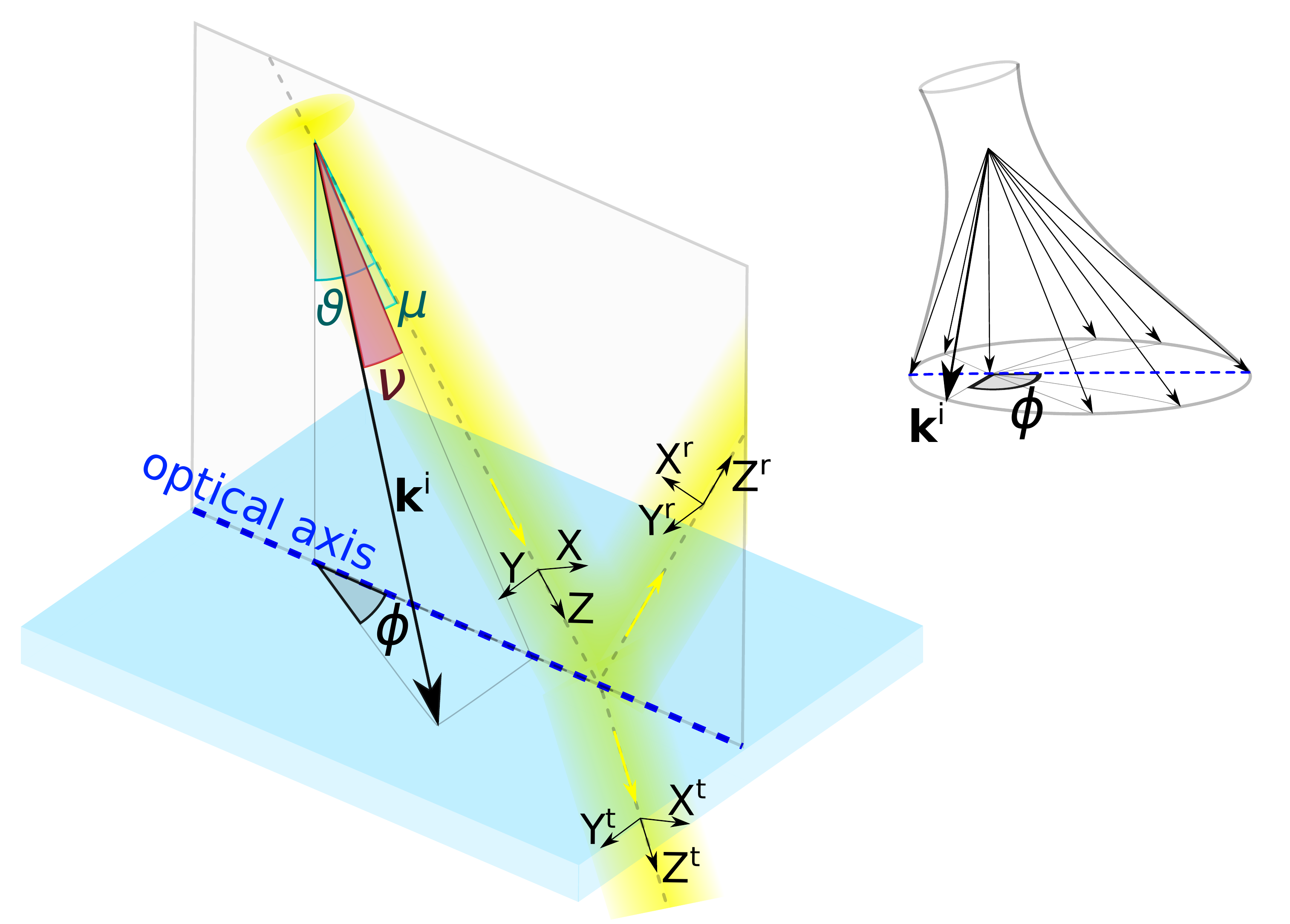}
	\caption{Schematic geometry of the problem. The long black arrow marks the non-central constituent plane wave with incident wavevector $\mathbf{k}^{i}$, angular in-plane ($\mu$) and out-of-plane ($\nu$) deflection components. Its plane of incidence forms a polar deflection angle $\phi$ with the one of a central plane wave. The yellow arrows are collinear with the wavevectors of the corresponding central plane waves. The incident angle of the central plane wave is marked as $\vartheta$. The inset schematically shows the incident wavevectors of the non-central plane waves of a beam forming the large polar deflection angles.}
	\label{Geometry}
	\vspace{-0pt}
\end{figure}

\textit{Standard theory of optical beam shifts}.--- 
We consider first the oblique incidence of a monochromatic optical beam on a flat vacuum-medium interface, see Fig.~\ref{Geometry}. 
%
%
%
%
%
%
The values of the shifts could be calculated within the Jones matrix formalism in paraxial approximation. The Jones matrix $\hat{T}^{a}$ relates the incident and reflected/transmitted plane wave amplitudes in the beam coordinate frame $|\mathbf{E}^a \rangle = \hat{T}^{a}(\vartheta, \mu, \nu) |\mathbf{E} \rangle$, where 
$|\mathbf{E} \rangle \propto |\mathbf{e} \rangle \cdot f(\mu,\nu)$ is the incident constituent plane-wave Jones vector, index $a = r,t$ denotes reflected and transmitted waves, respectively, $|\mathbf{e} \rangle = (e_X, e_Y)^{T} $ is the Jones vector of the central plane wave in the incident beam, and $ f(\mu,\nu) $ is the incident beam Fourier spectrum expressed through in-plane ($\mu$) and out-of-plane ($\nu$) deflections of the non-central wave vectors~\cite{bliokh2013goos}. 
%
%
%
The linear ($\langle X^a\rangle, \langle Y^a\rangle$) and angular ($\langle P_X^a\rangle, \langle P_Y^a\rangle$) GH ($\langle X^a\rangle, \langle P_X^a\rangle$) and IF ($\langle Y^a\rangle, \langle P_Y^a\rangle$) shifts are then obtained using their original definition meaning the quantum-mechanical expectation values of the coordinate and momentum operators~\cite{bliokh2013goos}:
\begin{equation}
\label{shifts_def}
\begin{split}
& \langle X^a \rangle, \langle Y^a \rangle =
\langle  {\mathbf{E}}^a \,|\,  i \partial / \partial k^a_{X,Y} \,|\, {\mathbf{E}}^a \rangle / N^a, \\
& \langle P_X^a \rangle, \langle P_Y^a \rangle =
\langle  {\mathbf{E}}^a \,|\, k^a_{X,Y} \,|\, {\mathbf{E}}^a \rangle / N^a,
\end{split}
\end{equation}
where the integration is taken over the transverse momentum components $k^a_X = \gamma^a k_0 \mu$ and $k^a_Y  = k_0 \nu$ in the beam coordinate frame, $k_0=2\pi/\lambda$ is the light wavevector in vacuum, $\lambda$ is the light wavelength, and factor $\{\gamma^{t,r}\}=\{1,-1\}$ accounts for the inversion of the $x$-components of the wavevectors with reflection. Here, $N^a = \langle {\mathbf{E}}^a | {\mathbf{E}}^a \rangle$ is the normalization factor proportional to the beam intensity. At large incidence angles, Jones matrix $\hat{T}^{a}$ is slightly inhomogeneous across the beam spectrum ($\partial \hat{T}^a / \partial k^a_{X,Y} \sim \lambda$), which leads to small shifts of the order of the light wavelength and angular spectrum variance for spatial and angular shifts, respectively (see Supplemental Material \cite{Supplement}).

\textit{Geometric Berry phase singularity}.--- 
%
%
%
%
The obvious absence of an IF shift under strictly normal incidence on one hand, and its singularity in approximate expressions for large incidence angles ($\propto \cot{\vartheta}$ where $\vartheta$ is the incidence angle~\cite{bliokh2013goos}) on another hand, assume the sharp IF shift peak in the vicinity of normal incidence. This motivates us to resolve explicitly this geometric singularity by analyzing the distribution of polar deflection angle~$\phi$, see Fig.~\ref{Geometry}. Under large incident angles the polar deflection angle $\phi$ of any constituent plane wave is small
\begin{equation}
    \phi = \frac{\nu}{\sin{\vartheta}} \ll 1. 
\end{equation}
Another situation takes place for the near-normal incidence, where different plane-wave components constituting the beam have planes of incidence oriented arbitrarily with respect to the central plane wave and cover a large range of polar deflection angles (at strictly normal incidence, $\phi \in [-\pi/2,\pi/2]$, see the inset of Fig.~\ref{Geometry}). In an anisotropic medium, the wide spreading of the incidence planes results in the significant difference of Fresnel's amplitudes for different plane-wave components. Thus, under some specific conditions (anisotropy degree, optical axis orientation, retardation phase of anisotropic slab), the optical beam incident at sufficiently small angles can experience anomalously large beam shifts. To show this explicitly, we note that at near-normal incidence, Jones matrix $\hat{T}^{a}(\vartheta, \mu, \nu)$ depends on incident angle $\vartheta$, in-plane $\mu$ and out-of-plane $\nu$ deflections only implicitly, through the polar deflection angle $\phi$ [$\hat{T}^{a}(\vartheta, \mu, \nu) \equiv \hat{T}^{a}(\phi)$], where the latter could be approximated as follows~\cite{Supplement}
\begin{equation}
    \phi(\vartheta,\mu,\nu) 
    \approx 
    \text{tan}^{-1}\left( \frac{\nu}{\vartheta + \mu} \right). 
    \label{phi}
\end{equation}
When $\vartheta$ is of the order of beam divergence, 
one can notice from 
Eq.~\eqref{phi} 
that the polar deflection angle $\phi$ may be large enough for a finite portion of the beam spectrum, and even close to $\pm90^\circ$ at $ \vartheta \simeq - \mu $. 
Thus, Jones matrix $\hat{T}^{a}(\phi)$ will be substantially inhomogeneous across the beam spectrum ($\partial \hat{T}^a / \partial k^a_{X,Y} \sim w_0$, where $w_0$ is the beam waist), which, in turn, will cause anomalous beam shifts (see Fig.~S2 in Supplemental Material~\cite{Supplement}).



\nocite{lekner1994optical}

In fact, the polar deflection angle at near-normal incidence ($\vartheta \rightarrow 0 $) is equivalent to the geometric Berry phase, $\Phi_B = -\phi \cos{\vartheta} \simeq -\phi$~\cite{bliokh2013goos}. Thus, the considered geometric singularity of the polar deflection angle simultaneously means the Berry phase singularity. The large IF shift arises from large transverse momentum derivatives of the Berry phase across the beam spectrum according to its definition. 

Another surprising consequence of Eq.~\eqref{phi} is that it leads to the anomalous GH shift. 
In conventional situation (large angles of incidence), the GH shifts are caused by the spatial dispersion of the scattering coefficients~\cite{bliokh2013goos}. 
In contrast, at near-normal incidence GH shift arises from large longitudinal momentum gradient of the Berry phase across the lateral parts of the beam spectrum. 
Note that such longitudinal gradients caused by strong spin-orbit coupling are negligible at large incident angles. 

\textit{Optical beam shifts at near-normal incidence}.---
To exhibit the near-normal Berry phase singularity we consider transmission through an anisotropic slab~-- a waveplate. Following the Jones matrix formalism we calculate analytically the GH and IF shifts of the Gaussian beam transmitted through the uniaxial waveplate when the beam incidence plane is parallel to the waveplate optical axis, which lies at the interface~(Fig.~\ref{Geometry}). The Gaussian beam angular spectrum is defined by the Fourier spectrum $f(\mu,\nu) = \exp[-\kappa^2 (\mu^2 + \nu^2)/2]$, where 
$\kappa=k_0 w_0/\sqrt{2}$ is the inverse beam divergence.
Results for the four transmitted beam shifts read (the detailed derivation is done in Supplemental Material~\cite{Supplement})
\begin{eqnarray}
\label{PX_uniax}
&&\langle \widetilde{P}_X \rangle
=
S_1 \cdot \tau_- \cdot \gamma^{t} \frac{\Lambda_X(\kappa \vartheta)}{\vartheta N^t},\\
\label{PY_uniax}
&&\langle \widetilde{P}_Y \rangle
=
S_2 \cdot \tau_- \cdot \frac{\Lambda_Y(\kappa \vartheta)}{ \vartheta N^t},\\
\label{X_uniax}
&&\langle \widetilde{X} \rangle =
- S_1 \cdot \tau_\times \cdot \gamma^{t} \frac{\Lambda_X(\kappa \vartheta)}{ \vartheta N^t},\\
\label{Y_uniax}
&&\langle \widetilde{Y} \rangle
=
- S_2 \cdot \tau_\times \cdot  \frac{\Lambda_Y(\kappa \vartheta)}{\vartheta N^t}
- 
S_3 \cdot 
|t_-|^2 
\frac{1 - e^{-\kappa ^2 \vartheta ^2}}{\vartheta N^t} 
,
\end{eqnarray}
where $ S_1=|e_x|^2 - |e_y|^2 $, $S_2 = 2 \Re[e_x^{*} e_y]$, and $S_3 = 2 \Im[e_x^{*} e_y]$ are the corresponding Stokes parameters of the incident beam, $ N^t = (\tau_+ + \tau_- \, S_1 \, \Lambda_Y(\kappa \vartheta) )/2$ is the squared norm of the transmitted beam, $ t_- = t_e - t_o $, $ \tau_{+,-} = |t_e|^2 \pm |t_o|^2$ , and $ \tau_\times = 2 \Im(t_e t_o^*) $ are real coefficients depending on the ordinary $ t_o $ and extraordinary $ t_e $ transmission amplitudes for the plane wave under normal incidence~\cite{Supplement}. Hereinafter, we use the dimensionless shifts values defined as $\langle \widetilde{P}_{X,Y} \rangle = \langle P_{X,Y} \rangle \cdot \kappa^2/k$ and $\langle \widetilde{X}, \widetilde{Y} \rangle = \langle {X}, {Y} \rangle \cdot k$. The nonlinear geometric resonant factors $\Lambda_Y$ and $\Lambda_X$ are the following
\begin{equation}
\begin{split}
\label{LQ}
& \Lambda_X(\kappa \vartheta)
=
\frac{
\left( 6 + 4 \kappa ^2 \vartheta ^2 + \kappa ^4 \vartheta ^4 \right) e^{-\kappa ^2 \vartheta ^2} + 2 \kappa ^2 \vartheta ^2 - 6}{\kappa ^4 \vartheta ^4},\\
& \Lambda_Y(\kappa \vartheta)
=
\frac{
6 - 4 \kappa ^2 \vartheta ^2 + \kappa ^4 \vartheta ^4 
- 2 e^{-\kappa ^2 \vartheta ^2} \left(\kappa ^2 \vartheta ^2+3\right)}{\kappa ^4 \vartheta ^4}.
\end{split}
\end{equation}
Note that the squared norm of transmitted beam state $ N^t(\vartheta) $ plays a role of a slight renormalization, in sharp contrast to conventional incident angles cases, where it causes singularity in the shifts by approaching zero~\cite{gotte2014eigenpolarizations,gotte2013limits,aiello2009brewster,merano2009observing,soboleva2012giant,salasnich2012enhancement,zhou2019controlling,hosten2008observation,gorodetski2012weak}. For the reflected beam, we arrive at the same results as in Eqs.~\eqref{PX_uniax}-\eqref{Y_uniax} with substitutions $\{t_e,t_o\} \rightarrow \{r_e,r_o\}$, $\gamma^{t} \rightarrow \gamma^{r}$, and $ N^t\rightarrow N^r$. 

Equations \eqref{PX_uniax}-\eqref{Y_uniax} are the main result of this work applicable for any cases except $\kappa \vartheta \gg 1$. One can see that each shift in Eqs.~\eqref{PX_uniax}-\eqref{Y_uniax} is the product of three factors -- incident beam polarization (Stokes parameters), material properties of the uniaxial slab ($\tau_{+,-,\times}$) and geometric resonance [$f(\kappa \vartheta)$]. The second term in the expression~\eqref{Y_uniax} for the spatial IF shift is a counterpart of the anomalous PSHE connected with the circular polarization of the incident beam. Moreover, due to the presence of the first term in~\eqref{Y_uniax} the IF shift can be obtained with the arbitrary polarization state except strictly TM- or TE-polarization.

We also note the remarkable connection of the four shifts to the geometric resonant terms $\Lambda_{X,Y}/(\vartheta N^t)$. 
First, there is a proportionality between spatial and angular shifts, which is only broken by the spin-Hall term in Eq.~\eqref{Y_uniax}. 
The simultaneous appearance of all possible optical beam shifts (namely, spatial and angular, GH and IF shifts) at the same near-normal angle of incidence is unique. At large incident angles ($\kappa \vartheta \gg 1$), spatial and angular GH shifts can be simultaneously observed only in special cases such as lossy media~\cite{aiello2009duality} or vortex beams~\cite{Bliokh09vortex}. Second, there is a further similarity of geometric resonant terms along $ Y $ and $ X $ directions, whereas $\Lambda_X(\kappa \vartheta) = e^{-\kappa^2 \vartheta^2} \Lambda_Y(i \kappa \vartheta)$. This reflects the fact that all anomalous near-normal incident shifts have the same origin, namely the geometric Berry phase singularity. In addition, the near-normal anomalous shifts form a connection to superoscillation-related physics~\cite{berry2019superosc}. 

\begin{figure}[t!]
	\includegraphics [width=0.40\textwidth]{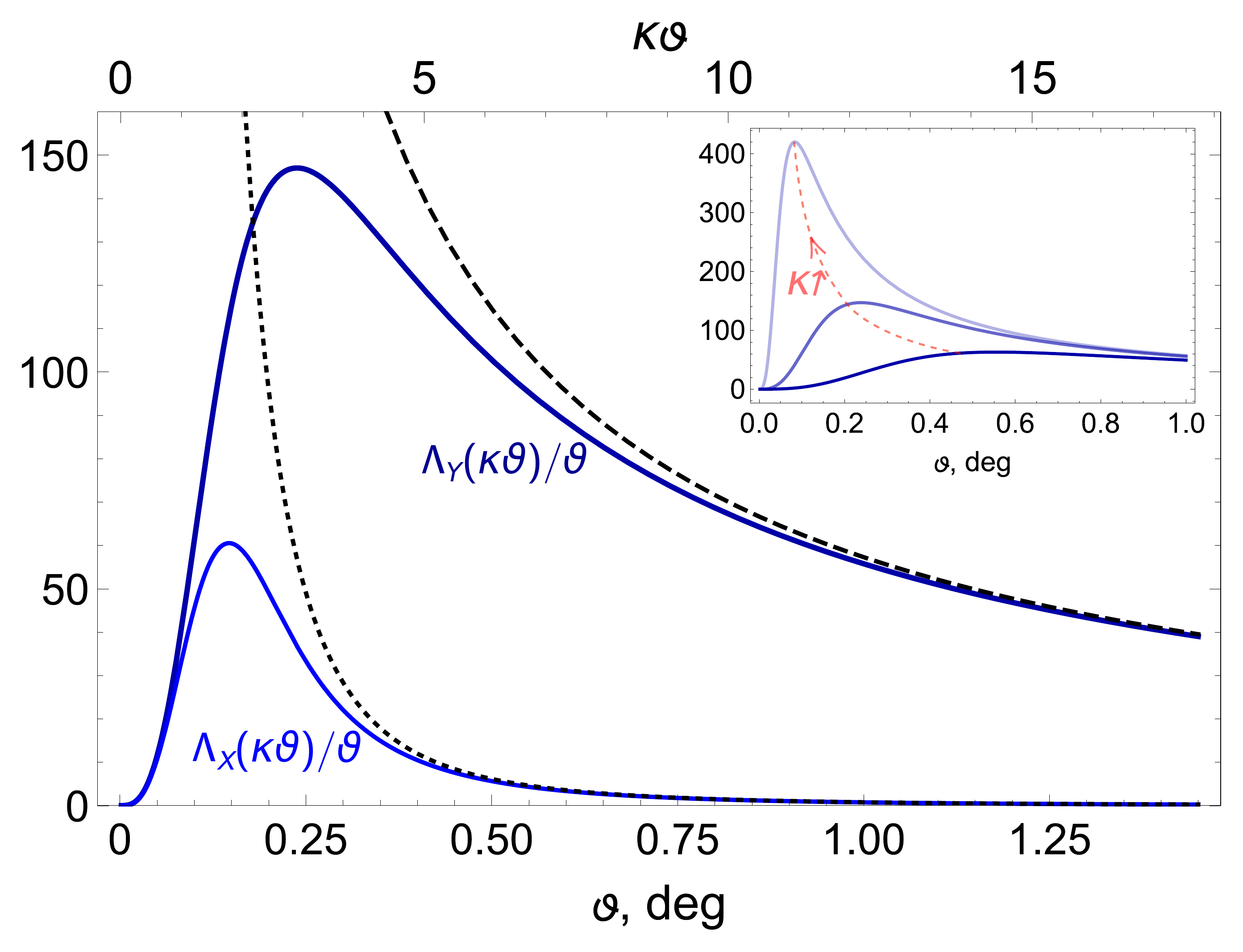}
	\caption{
		Nonlinear geometric resonant factors \eqref{LQ} appearing in shifts \eqref{PX_uniax}-\eqref{Y_uniax}: $ \Lambda_Y(\kappa \vartheta)/\vartheta $ (dark blue) and $ \Lambda_X(\kappa \vartheta)/\vartheta $ (light blue), along with their asymptotes at large incidence angles $ 1/\vartheta $ (dashed black) and $ 2/(\kappa^2 \vartheta^3) $ (dotted black), respectively. Here, we used the inverse beam divergence $ \kappa=700 $, namely $ \lambda=630\,\text{nm},\,w_0\simeq 100\,\mu \text{m} $. Inset shows the dependence of function $ \Lambda_Y(\vartheta)/\vartheta $ on the inverse beam divergence $ \kappa $. The arrow indicates the sequence of increasing beam waist, $ \kappa~=~300, \, 700, \, 2000 $, respectively.
	}\label{Plot_LQ}
	\vspace{-0pt}
\end{figure}

\textit{Impact of beam width}.---
The optical beam shifts under near-normal incidence  
depend substantially on the beam waist $w_0$ via the inverse beam divergence $\kappa$, as it is expressed by nonlinear functions $\Lambda_{X,Y}(\kappa \vartheta)$ [Eq.~\eqref{LQ}]. The IF shifts are anomalous and have asymptotes $ \propto 1/\vartheta $ at large incidence angles ($\kappa \vartheta \gg 1$) which is hinted by the standard theory~\cite{bliokh2013goos}. The GH shifts have higher-order asymptotes at large incident angles $ \propto 1/(\vartheta^3\kappa^2) $ [Fig.~\ref{Plot_LQ}]. This explains why the anomaly in GH shifts was not anticipated by the standard theory for uniaxial slab, in which the spatial GH shift is absent except the case of total internal reflection. Moreover, one can notice that the amplitude of the optical beam shifts peak is proportional to the beam waist, while the corresponding critical angle is inversely proportional to it (see the inset in Fig.~\ref{Plot_LQ}). In general, the critical angle corresponding to the maxima of GH and IF shifts is about a few tenths of a degree for a beam width of several dozen microns at $\lambda = 630$~nm. This result absolutely agrees with the recent experimental observations of the anomalous PSHE under near-normal incidence at hyperbolic metamaterials~\cite{takayama2018photonic,kim2019observation}. 


\textit{Dependence on slab material parameters}.---
From a rigorous analysis of near-normal shifts in isotropic media we found that the geometric Berry phase singularity at near-normal incidence is indeed manifested in the shifts only when the difference of Fresnel coefficients for TM- and TE-polarizations is finite at normal incidence~\cite{Supplement}. Then, we focus on the conventional waveplate with optical axis lying at the interface within a beam incidence plane. In the general case, the proposed generalized theory could be applied to arbitrary optical axis orientation and extended to biaxial and bianisotropic media. A particular case of an uniaxial slab with the optical axis perpendicular to the interface has been recently considered in Refs.~\cite{Zhu2021PSHE,Ling2021}, where the enhanced PSHE combined with the spin-controlled vortex generation was observed and explained by the divergence of the Berry phase.

\begin{figure}[t!]
	\includegraphics [width=0.33\textwidth]{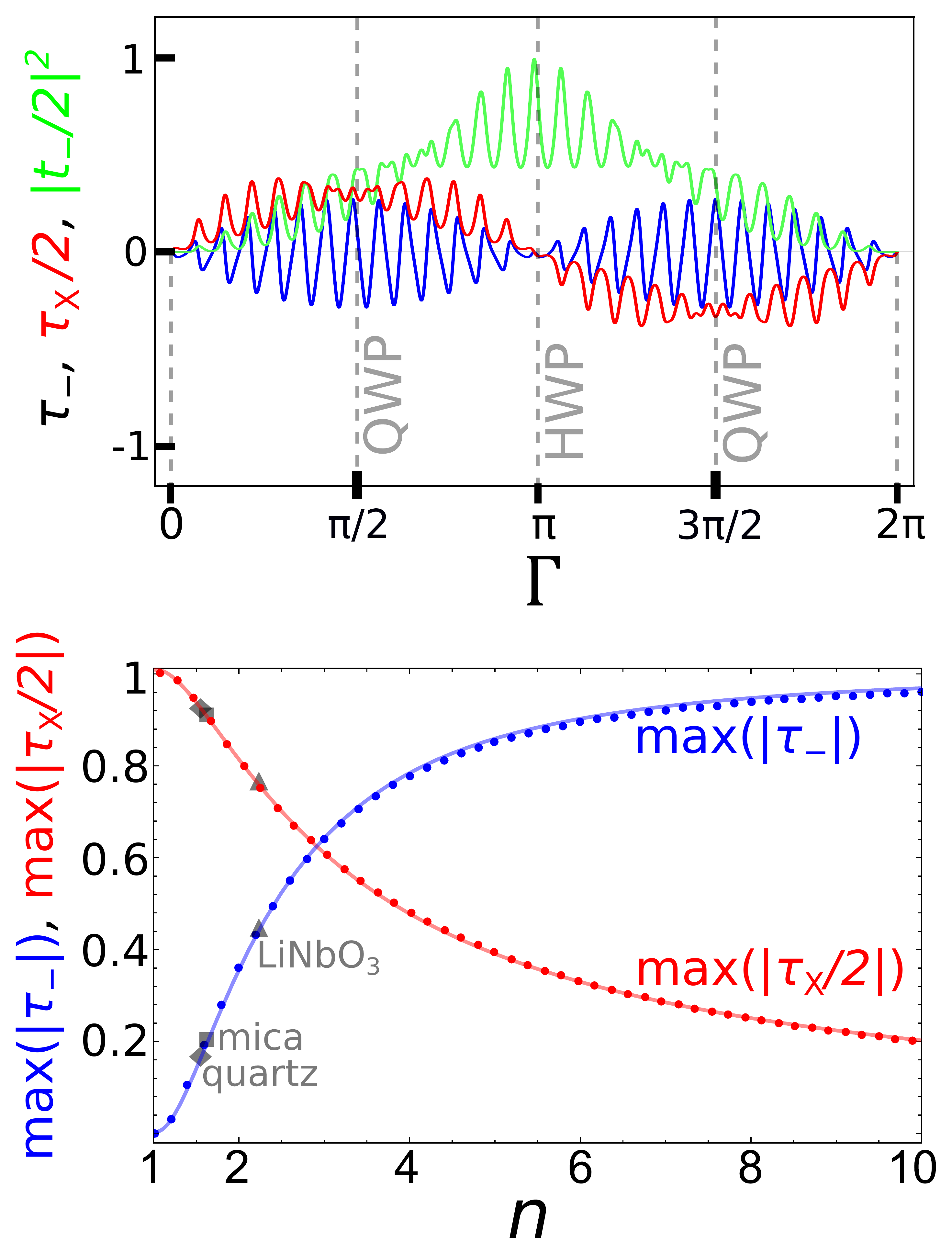}
	\caption{\textit{Upper panel}: $\tau_-$ (blue curve), $\tau_\times/2$ (red curve) and $|t_-|^2/2$ (green curve) as functions of waveplate retardation $\Gamma$ (for $n_o=2.6$, $n_e=2.8$). 
	\textit{Lower panel}: dependence of $\text{max}(|\tau_-|)$ and $\text{max}(|\tau_\times/2|)$ 
	on the average refractive index of the waveplate $n = (n_o + n_e)/2$ (points~-- numerics, solid curves~-- analytics, see Supplemental Material~\cite{Supplement}). 
	} 
	\label{JMoffdiag}
	\vspace{-0pt}
\end{figure}

All material dependences in Eqs.~\eqref{PX_uniax}-\eqref{Y_uniax} are factored out in four dimensionless coefficients, $ \tau_{+}, \tau_-, \tau_\times $, and $t_-$. Both angular shifts are proportional to $\tau_-$, while the spatial shifts are proportional to $\tau_\times$ except for the PSHE term in the spatial IF shift which is proportional to $|t_-|^2$. 
Therefore, we search numerically for the maximum values of these parameters as functions of the waveplate retardation $\Gamma = (n_e - n_o) k_0 \delta_z$, where $\delta_z$ is the slab thickness, and $n = (n_o + n_e)/2$ is the average refractive index of the uniaxial slab, see Fig.~\ref{JMoffdiag}. First, we find that both $|\tau_+|$ and $|\tau_\times|$ achieve maximum for the quarter-wave plate condition, i.e. for $\Gamma=\pm\pi/2 \,(\text{mod}\,\,2\pi)$, while $|t_-|^2/2$ reaches maximum value for a half-wave plate (HWP). This may also be shown analytically for small birefringence $|n_e - n_o|\ll~1$, see the Supplemental Material~\cite{Supplement}. The dependence of all parameters on $\Gamma$ is oscillatory due to the Fabry-P\'erot resonances in the slab~\cite{Supplement,luo2011enhanced}. 
Secondly, we consider a QWP and show that $\text{max}(|\tau_+|)$ grows as a function of average refractive index $n$ up to the saturation limit $\text{max}(|\tau_+|) \approx 1$ at $n \approx 10$, while $\text{max}(|\tau_\times/2|)$ reaches the same maximum, but in a distinct limit $n\rightarrow 1$. Hence, for realistic QWP with $n \approx 3 $, $|\tau_+|$ and $|\tau_\times|$ can be around 0.6 and 1.2, respectively. 
Thus the low-birefringence QWP is a perfect candidate for observing all four types of anomalous shifts simultaneously (although PSHE has maximum for a HWP).



\begin{figure}[t!]
	\centering
	\includegraphics [width=0.48\textwidth]{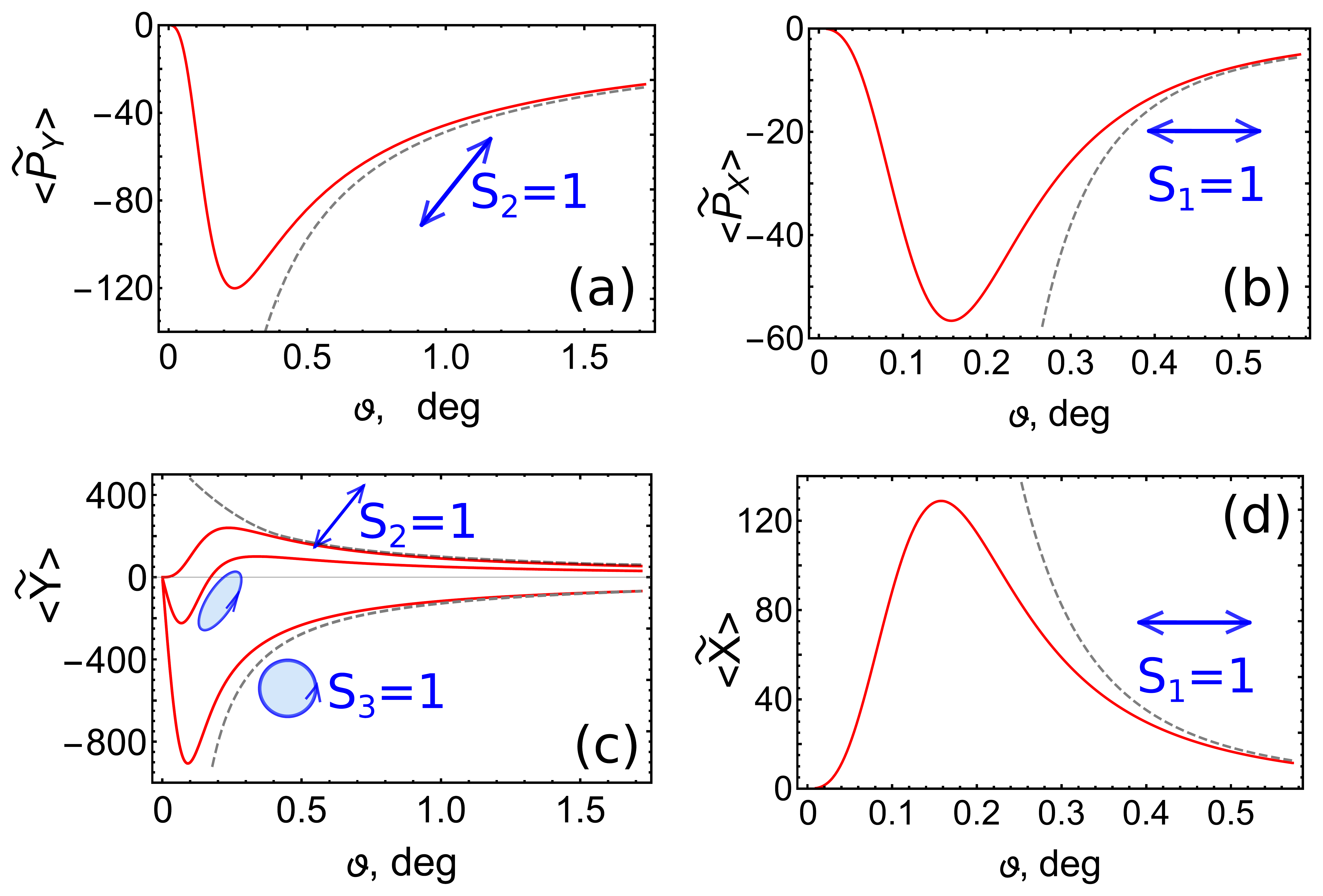}
	\caption{
		The resonant angular IF (a), angular GH (b), spatial IF (c), and spatial GH (d) shifts of the beam incident near-normally at a QWP (curve with one node in panel (c) corresponds to elliptic input polarization with Stokes parameters $S_2=0.95, \, S_3=0.31$).
		The gray dashed lines correspond to $ \propto 1/\vartheta $ (for IF) and $ \propto 1/(\kappa^2\vartheta^3) $ (for GH) asymptotes at large incidence angles ($\kappa \vartheta \gg 1$). 
		\textit{Parameters:} $n_o = 2.6, n_e = 2.7$, $\delta_z = 7.5 \lambda$ ($ \tau_-=-0.58 $, $ \tau_\times=-1.32 $, $ |\tau_-|^2=1.44 $, $ \tau_+ = 1.42 $). 
	}\label{XYsweeps}
	\vspace{-0pt}
\end{figure}

Fig.~\ref{XYsweeps} shows the optical beam shifts dependencies on the small incidence angle for the beam transmitted through the QWP with optical axis lying in the beam plane of incidence. It is important to note that a linear IF shift can change the shift direction to the opposite one at some angle under elliptically-polarized beam illumination (Fig.~\ref{XYsweeps}c). This is caused by the interplay between differently polarized terms of the linear IF shift [Eq.~\eqref{Y_uniax}]. 
The switching of the IF direction could be also observed for strongly anisotropic systems due to the polarization mixing, especially when the optical axis is arbitrarily oriented~\cite{mazanov2020photonic}.  

\begin{figure}[t!]
    \centering
	\includegraphics [width=0.9\textwidth]{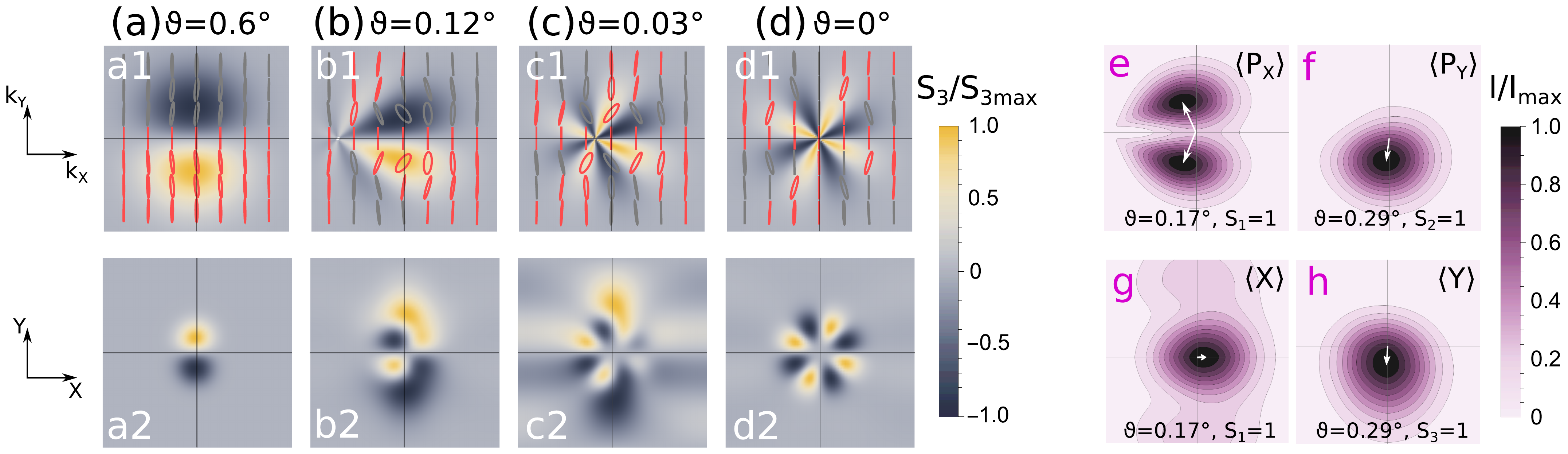}
	\caption{
		\emph{Panels} a-d: 
		normalized helicity (Stokes parameter $ S_3 $) patterns in $ \mathbf{k}$-space (a1-d1) and in real space (a2-d2) for the ordinary polarized ($ S_1=1 $) beam near-normally transmitted through QWP. The ellipses in panels (a1-d1) indicate the corresponding polarization profile in $\mathbf{k}$-space. 
		\emph{Panels} e-h: intensity profiles at maximum linear and angular GH and IF shifts (shown by white arrows) of transmitted beam for their respective eigenpolarizations. 
		Parameters are the same as in Fig.~\ref{XYsweeps}. 
		The plots in the panels are the squares with the side $10 w_0$ for all real-space panels, and $10 / \kappa$ for all $ \mathbf{k}$-space panels. 
	}\label{eigen_effects_meta}
\end{figure}

\textit{Intensity and helicity profiles of the anomalously shifted beam}.---
The  intensity and $S_3$ (helicity) patterns simulated with Jones matrix $ \hat{T}^t $ 
are shown in Fig.~\ref{eigen_effects_meta} for the corresponding input eigenpolarizations. 
The helicity dynamics for an ordinary TM-polarized beam  ($S_1=1$) provides additional insight into the nature of the anomalous shifts at near-normal incidence. When the incidence angle $\vartheta$ exceeds the beam divergence $1/\kappa$, PSHE naturally occurs owing to the so-called circular birefringence (Fig.~\ref{eigen_effects_meta}a)~\cite{bliokh2016spin}. 
At near-normal incidence $ \vartheta \lesssim 1/\kappa $, spin-Hall doublets from lateral and opposite parts of beam Fourier spectrum become prominent, and an octupole helicity profile is built up in $\textbf{k}$-space with a phase singularity at a distance $ \sim k_0\vartheta $ from the geometric-optics beam axis (Figs.~\ref{eigen_effects_meta}b-\ref{eigen_effects_meta}c). 
In real space, at $ \vartheta=0 $, the helicity is carried by a mode resembling a Bessel-Gaussian mode with $ m=4 $~\cite{ito2010generation} and intensity null
on the beam axis (Fig.~\ref{eigen_effects_meta}d). 
This mode comes from the extraordinary component induced by the waveplate and deforms along with the $ \mathbf{k}$-space profile, accompanying the anomalous spatial shifts. Transverse helicity imbalance accompanies the angular IF shift for incident beam polarization different from $S_1$.
The presence of such a mode 
and the polarization singularity provide another close connection to the superoscillations~\cite{berry2019superosc}. Finally, we demonstrate explicitly the beam shifts of all four kinds in Figs.~\ref{eigen_effects_meta}e-\ref{eigen_effects_meta}h. The anomalous GH angular shifts are accompanied by  longitudinal beam spectrum deformations around this polarization singularity (Fig.~\ref{eigen_effects_meta}e). 


To conclude, we have developed a theory of spatial and angular shifts of optical beams undergoing near-normal incidence. The theory is able to explain experimental peculiarities of strongly enhanced IF shifts obtained when the angle of incidence was less than one degree, and predicts the additional anomalous GH shifts. The principal point of the theory is that anomalously large shifts of all types are shown to have the same fundamental origin~-- the singularity in the Berry phase appearing in the beam Fourier spectrum. The results have been illustrated for an analytically treated example of a uniaxial waveplate. We propose a very simple configuration with a quarter-waveplate, which supports anomalously large simultaneous near-normal incidence shifts of all kinds in manifold times exceeding shifts observed separately in conventional cases. Thus, the anomalous near-normal shifts are completely feasible for direct experimental verification. 
The results obtained could arise for other types of waves (acoustic, electron beams, etc.) material anisotropy such as biaxial and bianisotropic media. 

O.~Y. and A.~B. acknowledge support from the Foundation for the Advancement of Theoretical Physics and Mathematics "BASIS". The authors thank Konstantin Bliokh from RIKEN for fruitful discussions.

\onecolumngrid

\twocolumngrid

\bibliography{refs}

\end{document}


\title{Supplemental Material: \\ On anomalous optical beam shifts at near-normal incidence}
	
\author{Maxim Mazanov}
\email{mazanovmax@gmail.com}
\affiliation{V. N. Karazin Kharkiv National University, Kharkiv, Ukraine}
\affiliation{School of Physics and Engineering, ITMO University, St. Petersburg, Russia}
\author{Oleh Yermakov}
\affiliation{V. N. Karazin Kharkiv National University, Kharkiv, Ukraine}
\affiliation{School of Physics and Engineering, ITMO University, St. Petersburg, Russia}
\author{Andrey Bogdanov}
\affiliation{School of Physics and Engineering, ITMO University, St. Petersburg, Russia}
\author{Andrei Lavrinenko}
\affiliation{Department of Photonics Engineering, Technical University of Denmark, Kgs. Lyngby, Denmark}


\maketitle

\widetext
\setcounter{equation}{0}
\setcounter{figure}{0}
\setcounter{table}{0}
\setcounter{page}{1}
\setcounter{section}{0}
\makeatletter
\renewcommand{\theequation}{S\arabic{equation}}
\renewcommand{\thefigure}{S\arabic{figure}}
\renewcommand{\bibnumfmt}[1]{[S#1]}
\renewcommand{\citenumfont}[1]{S#1}

\section{Supplemental Note 1: Polar deflection angle and geometric Berry phase}

\begin{figure}[h!]
	\includegraphics [width=0.33\textwidth]{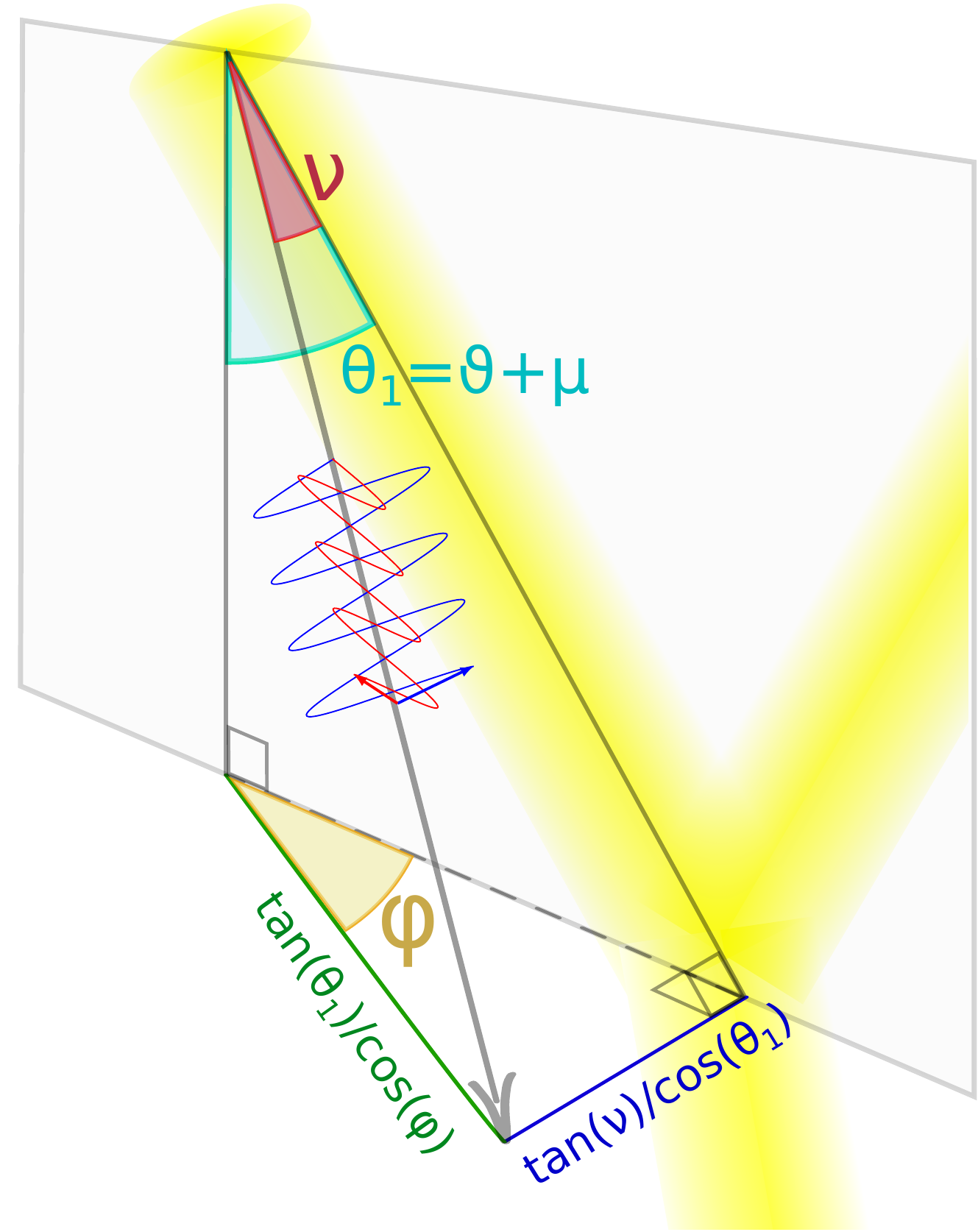}
	\caption{Schematic geometry for the calculation of polar deflection angle $\phi$ for a component plane wave with wave vector depicted by a gray arrow. The gray vertical plane marks the incidence plane of central plane-wave in the beam.}
	\label{phi_Geom}
\end{figure}

In order to demonstrate the keynote point distinguishing our work from all other studies devoted to GH and IF shifts, we derive the strict expression for the polar deflection angle $\phi$ [Eq.~(3)], responsible for the deflection of non-central constituent plane wave from the central one. The corresponding relation between the polar deflection angle $\phi$ and the incident angle $\vartheta$, small in-plane $\mu$ and out-of-plane $\nu$ deflections can be derived from Fig.~\ref{phi_Geom}:
\begin{equation}
\label{tan phi}
\tan\phi = \frac{\tan\nu}{ \sin\theta}
\approx
\frac{\nu}{\vartheta + \mu}. \\
\end{equation}
where $\theta = \vartheta + \mu$ is the total incident angle of the non-central wave in the plane of incidence. Importantly, the explicit form of $\sin{\phi}$ and $\cos{\phi}$ could be expressed as follows:
\begin{equation}
\begin{split}
\label{cos-sin phi}
& \sin\phi \approx
\frac{\nu/\theta}{\sqrt{(\nu/\theta)^2+1} } = \frac{\nu \cdot \text{sign}(\vartheta + \mu)}{\sqrt{\nu^2 + (\vartheta + \mu)^2}}, \\
& \cos\phi \approx
\frac{1}{\sqrt{(\nu/\theta)^2+1}} = \frac{|\vartheta + \mu|}{\sqrt{\nu^2 + (\vartheta + \mu)^2}}.
\end{split}
\end{equation} 

Within the previous studies, e.g.~\cite{bliokh2013goos}, it was assumed that $\nu,\mu \ll \vartheta$, so the polar deflection angle was determined as $\phi = \nu / \sin{\vartheta}$ being always negligibly small in comparison to the angle of incidence $\phi \ll \vartheta$. This statement is absolutely relevant for any not near-zero angle~\cite{bliokh2013goos}.

However, in the case of near-normal incidence, when $\vartheta$ becomes comparable with $\nu$ and $\mu$, one can notice from Eq.~\eqref{tan phi} that the polar deflection angle may be extremely large. Under condition $\vartheta + \mu = 0$ it could be even close to $90^\circ$. 
Another appealing feature of Eq.~\eqref{tan phi} is the fact that $\phi$ becomes $\mu$-dependent under near-normal incidence. This arbitrary beam-width-dependent polar angle deflection under the special case of near-normal incidence is the keynote fundamental insight of this work leading to the unconventional behaviour of GH and IF shifts described in the main text. In particular, beam waist can strongly affect not only the amplitude but also the sign of $\phi$. Therefore, the planes of incidence for different plane waves constituting the beam can be substantially different.

Under near-normal incident angles, the geometric Berry phase of constituent non-central plane wave is induced by the azimuthal rotation of the plane of incidence and is completely determined by the polar deflection angle $\Phi_B = -\phi \cos{\vartheta} \simeq -\phi$~\cite{bliokh2013goos}. So, the intriguing properties of the optical beam shifts under near-normal incidence are defined by the singularity of the geometric Berry phase.

\section{Supplemental Note 2: Jones matrix formalism for optical beam incident on uniaxial slab along optical axis at near-normal incidence}

We follow the formalism of the Jones matrix describing the beam interaction with the interface including spatial dispersion effects in the momentum space representation~\cite{bliokh2013goos}
\begin{equation}
\label{T-matrix_formalism}
\hat{T}^{a} = \hat{U}^{a\dag}(\vartheta^{a},\mathbf{k}^{a}) \, \hat{F}^{a}(\vartheta,\mathbf{k}^a) \, \hat{U}^{a}(\vartheta^{a},\mathbf{k}^{a}).
\end{equation}
Here, index $a$ denotes reflected ($r$) and transmitted ($t$) beams, respectively; $\hat{U}^a(\vartheta^a,\mathbf{k}^a)$ is the rotational transformation of the fields from the beam coordinate frame to the spherical basis of the TE and TM modes; $\hat{F}^a(\vartheta,\mathbf{k}) = (f^a_{pp},f^a_{ps};f^a_{sp},f^a_{ss})$ is the Fresnel matrix for the fields in the spherical basis of the TE and TM modes; $\vartheta$ is the angle of incidence, $\vartheta^a$ is the reflected/refracted angle; $\mathbf{k}^a \simeq \mathbf{k}^a_c + (k^a_X \mathbf{u}^a_X + k^a_Y \mathbf{u}_Y) $ is the wave vector of constituent plane wave of the beam, $\mathbf{k}^a_c = n^a k_0 (\sin{\vartheta^a}, 0, \cos{\vartheta^a})$ is the wave vector of the beam central plane wave, $n^a$ is the refractive index of the corresponding medium, $k_0 = 2 \pi / \lambda$ is the free-space wave vector, $\lambda$ is the wavelength of incident light, $\mathbf{u}^a_X$ and $\mathbf{u}_Y$ are the unit vectors in the geometric-optics beam coordinate system (beam propagates along $z$-axis), $k^a_X = k_0 \gamma^a \mu$ and $k^a_Y = k_0 \nu$ are the transverse wave vector components of the constituent plane waves, $\gamma^a = \cos{\vartheta} / \cos{\vartheta^a}$, $\mu$ and $\nu$ are the in-plane and out-of-plane deflections of non-central wave vectors as it is shown in Fig.~\ref{phi_Geom}. Hereinafter, we consider only square $2\times2$ matrices corresponding to the transverse components of the fields which are essential for a paraxial beam. Next, we derive the rotation matrix $ \hat{U}^a(\vartheta,\mathbf{k}) $. As already mentioned, this matrix corresponds to the rotational transformation of the electric field of the particular constituent plane wave in a beam from the beam coordinate frame to the (local) spherical basis of the TE and TM modes, where the Fresnel matrix would be diagonal, which involves three consequent rotations $\hat{U}^a\left(\vartheta^{a}, \mathbf{k}^{a}\right)=\hat{R}_{y}\left(\theta^{a}\right) \hat{R}_{z}\left(\phi^{a}\right) \hat{R}_{y}\left(-\vartheta^{a}\right)$~\cite{bliokh2013goos}. Here, the first (in acting order) is the rotation from beam coordinate frame to laboratory frame $\hat{R}_{y}\left(-\vartheta^{a}\right)$, followed by rotation from the laboratory frame to the local spherical coordinate frame (which is aligned with the constituent plane-wave component plane of incidence and has axis $z'$ normal to the interface), obtained with rotation by a polar deflection angle $\hat{R}_{z}\left(\phi^{a}\right)$, and the rotation $\hat{R}_{y}\left(\theta^{a}\right)$ back to the beam coordinate frame. The part of matrix $\hat{U}^a(\vartheta,\mathbf{k})$ for transverse fields has the following form:

\begin{equation}
\label{U0}
\hat{U}^a(\vartheta,\mathbf{k})
=
\left(
\begin{array}{cc}
\cos \vartheta \cos \theta \cos \phi+\sin \vartheta \sin \theta & \cos \theta \sin \phi \\
-\cos \vartheta \sin \phi & \cos \phi \\
\end{array}
\right)
.
\end{equation}
Under near-normal incidence, we may set in Eq.~\eqref{U0} that $ \cos \vartheta \approx 1 $, $ \cos \theta \approx 1 $ in the leading order, and neglect $ \sin \vartheta \sin \theta \approx \vartheta \sqrt{(\nu)^2 + (\vartheta + \mu)^2} \ll  \cos \phi $, leaving $ \mu $-dependence only in $ \sin \phi $ and $ \cos \phi $ (which leads to neglecting terms $ \propto\mu^2 $ and higher-order terms in $ \vartheta $). 
Therefore, the rotation matrix in the leading-term approximation is just the matrix of rotation by polar angle $\phi$ in transverse beam coordinates:
\begin{equation}
\label{U}
\hat{U}^a(\phi)
\simeq
\left(
\begin{array}{cc}
\cos \phi & \sin \phi \\
-\sin \phi & \cos \phi \\
\end{array}
\right)
.
\end{equation}

The ordinary ($o$) and extraordinary ($e$) transmission and reflection coefficients for an uniaxial plate could be deduced using $ 4\times4 $ transfer matrix method~\cite{lekner1994optical}
\begin{eqnarray}
&& \label{t_eo}
t_{o,e}= \frac{
\left(1-\eta_{o,e}^2\right) \exp [i \tilde{\delta}_z n_{o,e}]
}{
1-\eta_{o,e}^2 \exp [2 i \tilde{\delta}_z n_{o,e}] }, \\
&& \label{r_eo}
r_{o,e}= \frac{
\eta_{o,e} \left(1-\exp [2 i \tilde{\delta}_z n_{o,e}]\right)
}{
1-\eta_{o,e}^2 \exp [2 i \tilde{\delta}_z n_{o,e}]
},
\end{eqnarray}
where $ n_o $ and $ n_e $ are the refractive indices for ordinary and extraordinary wave in an uniaxial crystal, respectively, $ \eta_{o,e} = (1-n_{o,e}) / (1+n_{o,e}) $, and $ \tilde{\delta}_z = k_0 \delta_z$ is the dimensionless normalized slab thickness, $\delta_z$ is the slab thickness. Here, we have neglected $ \mathcal{O}(\vartheta^2)$ terms in all transmission coefficients since $ \vartheta \rightarrow 0 $. 

In the general form, Fresnel matrices for transmitted ($\hat{F}^{t}$) and reflected ($\hat{F}^{r}$) plane waves can be expressed as follows:
\begin{equation}
\label{Fresnel}
    \hat{F}^{t} = \begin{pmatrix}
    t_{pp} & t_{ps}\\
    t_{sp} & t_{ss}
    \end{pmatrix}, \;
    \hat{F}^{r} = \begin{pmatrix}
    r_{pp} & r_{ps}\\
    r_{sp} & r_{ss}
    \end{pmatrix},
\end{equation}
where $t_{ij}$ and $r_{ij}$ are the complex transmission and reflection coefficients. The corresponding transmittance and reflectance are defined as $T_{ij} = |t_{ij}|^2$ and $R_{ij} = |r_{ij}|^2$, respectively. Indices $p$ and $s$ correspond to TM and TE polarization, respectively. The components of the transmissive Fresnel matrix~\eqref{Fresnel} are
\begin{equation}
\begin{split}
& t_{pp}=t_e \cos^2{\varphi} + t_o \sin^2{\varphi}, \\
& t_{ss}=t_o \cos^2{\varphi} + t_e \sin^2{\varphi}, \\
& t_{sp}=t_{ps}=(t_e-t_o) \sin{\varphi} \cos{\varphi},
\label{t_uniax}
\end{split}
\end{equation}
where $ \varphi $ is the angle between the incidence plane for constituent beam plane wave and the optical axis of an uniaxial crystal. The expression \eqref{t_uniax} shows that at nonzero angles $ \varphi $ off-diagonal components $ t_{sp}=t_{ps} $ of the Fresnel matrix mix $ s $ and $ p $ polarizations of a plane wave, with maximum mixing at $ \varphi = \pi/4 $.

Hereinafter, we consider the special case, when the optical axis of an uniaxial crystal lies at the intersection of the planes of the interface and central beam plane wave, so that $\phi \equiv \varphi$. Therefore, in this approximation $\vartheta$ enters $ \hat{F}^{t,r}(\phi) $ only implicitly through the deflection angle $ \phi $. This is the least polarization-mixing and counter-intuitive case for polarization dependence of beam shifts, while one can generally consider the arbitrary values of $\varphi$ leading to the increase of the polarization mixing effects. Moreover, this case also describes the optical axis lying within the interface of uniaxial crystal, but oriented perpendicularly to the central plane wave of a beam, with the corresponding substitution $t_{e,o} \to t_{o,e}$.

The Jones matrices for transmitted light in an uniaxial waveplate are readily derived from their definition in the main text, and at any plate thickness we find the following compact form for them:
\begin{equation}
\label{Tt_Jones}
\hat{T}^t
= \frac{1}{2}
\left(
\begin{array}{cc}
t^{+} + t^{-} \cos (4 \phi ) & t^{-} \sin (4 \phi ) \\
t^{-} \sin (4 \phi ) & t^{+} - t^{-} \cos (4 \phi ) \\
\end{array}
\right)
,
\end{equation}
where complex coefficients $ \tau^{+,-} $ and $ \rho^{+,-} $ are defined as follows 
\begin{equation}
\label{ab}
t^{+,-} = t_e \pm t_o.
\end{equation}
The product of Jones matrices appearing in Eq.~(1) could be expressed as follows
\begin{eqnarray}
	\label{TT}
	\hat{T}^{t\dag}	\hat{T}^{t} = \frac{1}{2}	\left(
	\begin{array}{cc}
	\tau^+ + \tau^- \cos (4 \phi ) & \tau^- \sin (4 \phi ) \\
	\tau^- \sin (4 \phi ) & \tau^+ - \tau^- \cos (4 \phi ) \\
	\end{array}
	\right), 
\end{eqnarray}
where
\begin{equation}
\label{xi-psi}
\tau^{+,-} = |t_e|^2 \pm |t_o|^2
.
\end{equation}

In the case of reflection, the product of Jones matrices~\eqref{TT} remains the same with proper substitution $t_{o,e} \to r_{o,e}$.

\section{Supplemental Note 3: Optical beam shifts for the quarter-wave plate under near-normal incidence}

In this section, we derive the explicit form of Jones matrices for a uniaxial waveplate and analyze the impact of materials parameters (namely, anisotropic refractive index tensor and thickness) of uniaxial slab on the magnitude of optical beam shifts. 

According to Eqs.~(4)-(7) of the main manuscript, we find that the angular and linear optical beam shifts are proportional to $\tau_-$ and $\tau_\times$, respectively. Here, we perform the analytical analysis for low-birefringent ($|n_o-n_e| \ll 1$) uniaxial plate in order to determine the waveplate retardation corresponding to maximum of $\tau_-$ and $\tau_\times$. In the case of transmitted beam, one can write
\begin{eqnarray}
\label{xi_max_analytical}
\label{2xi}
\tau_- 
\simeq 
x^2 \cdot \left( 
\frac{1}{|1-y|^2} - \frac{1}{|1-y\cdot e^{2i \Gamma}|^2}
\right)
,
\end{eqnarray}
where $\Gamma = (n_o - n_e) \tilde{\delta}_z$ is the waveplate retardation, $x = (1-\eta_e^2)$, $y = \eta_e^2 \cdot e^{2i\tilde{\delta}_z n_e}$. Fabri-P\'erot condition for constructive interference in a waveplate implies that the phase $2\tilde{\delta}_z n_e \equiv 0\,\mod \, 2\pi$, which ensures that $y$ should be a real number. In addition, since $|\eta_{e,o}| < 1$ we can state that $0 < (1-y) < 1$. Then, the maximum is achieved for the most different denominators in Eq.~\eqref{2xi}, i.e. for $e^{2i \Gamma} = -1$, which gives finally $\Gamma_{max} \equiv \pm \pi/2\,\mod \, 2\pi$ corresponding indeed to the quarter-wave plate retardation. 

Similarly, we can prove that maximum of $\tau_\times = 2\Im(t_e t_o^*)$ also appears for the quarter-wave retardation. First, we can write explicitly 
\begin{eqnarray}
\label{zeta_max_analytical}
\label{2zeta}
\zeta 
\simeq 
2 x^2 \cdot \Im \left( 
\frac{e^{-i \Gamma}}{(1-y) (1-y^* \cdot e^{2i \Gamma})}
\right)
.
\end{eqnarray}
Then, after noting that $y$ should be real-valued (see discussion for $\tau_-$ above) it takes the form 
\begin{eqnarray}
\label{xi_max_analytical}
\label{zeta_final}
\tau_\times 
\simeq 
- \frac{2 x^2 ( \sin\Gamma - y \sin3\Gamma)}{( 1-y)( 1+y^2 - 2 y \cos2\Gamma)}
. 
\end{eqnarray}
Therefore, maximum $\tau_\times$ is achieved for $\Gamma_{max} \equiv \pm \pi/2\,\mod \, 2\pi$, which is again the quarter-wave plate retardation. 

Analytic formulas for maximum values of $|\tau_-|$ and $|\tau_\times|$ at near-quarter-wave plate retardation thus read 
\begin{eqnarray}
\label{xi_zeta_max_analytical}
|\tau_+|_{max} 
\simeq 
\frac{4 \eta^2}{\left( 1 + \eta^2 \right)^2
}
, \qquad
|\tau_\times|_{max} 
\simeq 
\frac{2(1-\eta^2)}{1+\eta^2}
, 
\end{eqnarray}
where $ \eta = (1-n)(1+n) $, and $ n =(n_o + n_e) / 2 $ is the average refractive index. Analytical results~\eqref{xi_zeta_max_analytical} are plotted in the lower panel of Fig.~3 in the main text (solid lines) along with the numerical results (points), where we find a good agreement.

\begin{figure}[H]
    \centering
	\includegraphics [width=0.8\textwidth]{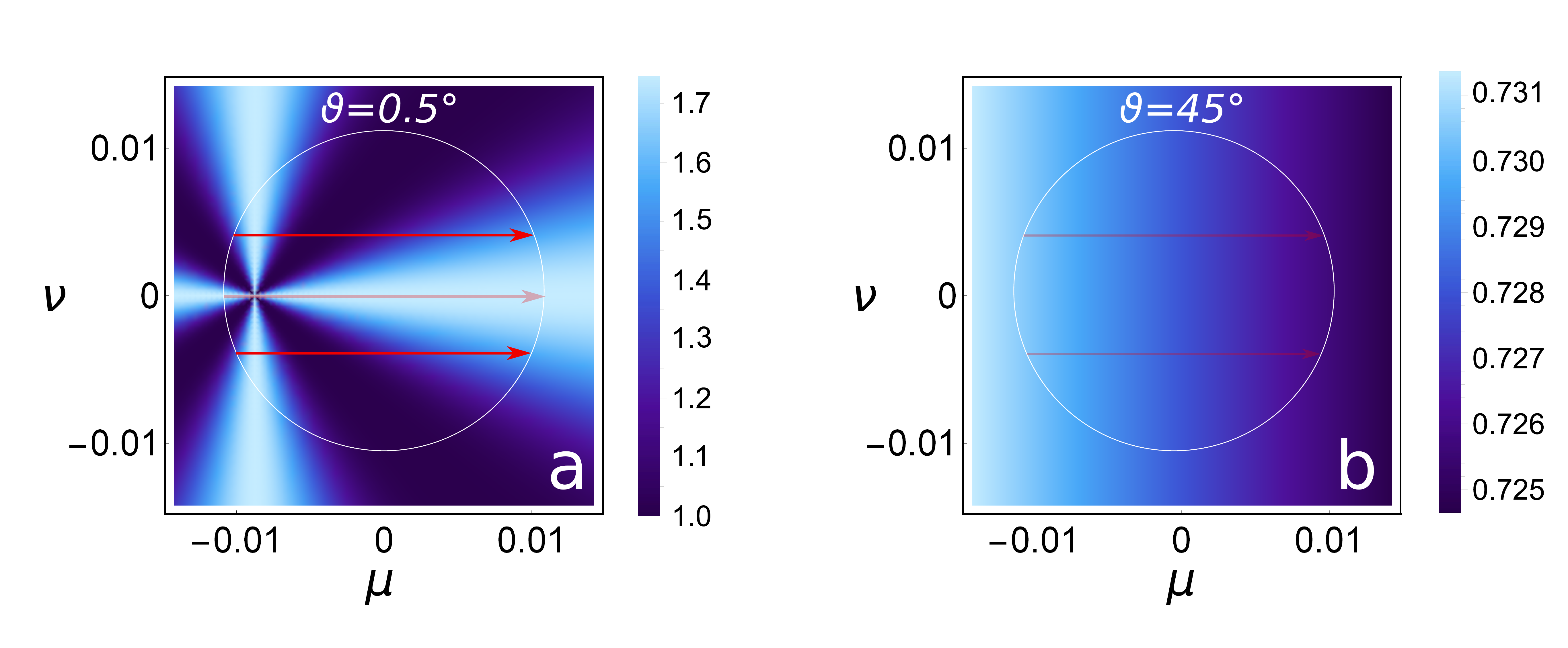}
	\caption{
        The absolute value of main-diagonal Jones matrix element $|\hat{T}^{t}_{11}|$ for (a) near-normal and (b) not-near-normal incidence in constituent plane wave deflection coordinates $\mu$ and $\nu$ (in radians). The white circle represents the beam angular width in $\mathbf{k}$-space. The brightness of the arrows schematically indicates the value of derivative, namely, pale and bright arrows correspond to $\partial \hat{T}^{t}_{11} / \partial \mu \ll \kappa$ and $\partial \hat{T}^{t}_{11} / \partial \mu  \sim \kappa$, respectively. Parameters: $n_o = 1.5,\, n_e = 1.7,\, \kappa=70,\, \tilde{\delta}_z = (\pi/2)/(n_e - n_o) $ (zero-order QWP). 
	}\label{GH_arg}
\end{figure}

Moreover, taking into account the form of Jones matrix \eqref{Tt_Jones}, we may see why anomalous GH shifts arise in the case of normal incidence at the QWP. It is well known that the GH shifts are proportional to the $\mu$-derivatives of Jones matrix elements $\hat{T}^{t}_{11}$ and $\hat{T}^{t}_{22}$~\cite{bliokh2013goos}. In Fig.~\ref{GH_arg}, we plot $|\hat{T}^{t}_{11}|$ for near-normal (left panel) and not-near-normal (right panel) incidence in constituent plane wave deflection coordinates $\mu$ and $\nu$. We see that at not-near-normal incidence, the derivative $\partial \hat{T}^{t}_{11} / \partial \mu \ll \kappa$ within the beam angular width and does not depend on $\nu$ to the first order. However, at near-normal incidence, the derivative $\partial \hat{T}^{t}_{11} / \partial \mu \sim \kappa$ for wavevectors with $\nu \neq 0$. Thus, the contribution from the \textit{lateral} plane waves is indeed the source of anomalous GH shifts at near-normal incidence.

\newpage

\section{Supplemental Note 4: Near-normal shifts in isotropic media}
In this section, we prove that the near-normal shifts in isotropic homogeneous media do not exhibit anomalous shifts of any kind. We find that all shifts are factorized into two parts: (i) a resonant geometric part and (ii) material- and polarization-dependent part, where the latter part destroys the geometric resonance. The only exception is the angular IF shift which has a maximum for strictly circular input polarization, which is however subwavelength and thus not anomalous. These results further suggest that a finite difference of Fresnel amplitudes for TE and TM waves at normal incidence (as it happens for any anisotropic medium, including low-birefringent uniaxial crystals) would be sufficient condition for the observation of the anomalous shifts. In this case, the geometric resonance term overcomes the material-dependent one and plays a crucial role for the observed optical beam shifts peaks.

Following Eq.~(3.13) from Ref.~\cite{bliokh2013goos} we write the Fresnel matrix for a homogeneous isotropic medium under small incidence angles $ \vartheta $ :
\begin{eqnarray}
\label{Fa}
\hat{F}^a(\vartheta,\mathbf{k})
=
\left(
\begin{array}{cc}
f^a_p (1+\mu^2 \tilde{\mathcal{X}}^a_p)
& 0 \\
0 & f^a_s (1+\mu^2 \tilde{\mathcal{X}}^a_s)  \\
\end{array}
\right)
.
\end{eqnarray}
Here $ \tilde{\mathcal{X}}^a_{p,s} = (1/f^a_{p,s}) \partial^2 f^a_{p,s} / \partial \vartheta^2 $ are the normalized second-order derivatives of Fresnel coefficients.

Then we derive the Jones matrix relating the incident and secondary fields in the beam coordinate frames according to Eq.~\eqref{T-matrix_formalism}:
\begin{equation}
	\label{Taperp}
	\hat{T}^a
	\simeq
	\left(
	\begin{array}{cc}
	f^a_p (1+\mu^2 \tilde{\mathcal{X}}^a_p) \cos^2\phi +
	f^a_s (1+\mu^2 \tilde{\mathcal{X}}^a_s) \sin^2\phi 
	& f^a_p \tilde{\mathcal{Y}}^a_p \sin\phi \cos\phi \\
	-f^a_s \tilde{\mathcal{Y}}^a_s \sin\phi \cos\phi & f^a_s (1+\mu^2 \tilde{\mathcal{X}}^a_s) \cos^2\phi +
	f^a_p (1+\mu^2 \tilde{\mathcal{X}}^a_p) \sin^2\phi  \\
	\end{array}
	\right),
\end{equation}
where the ``relative Fresnel coefficients` splittings'' $ \tilde{\mathcal{Y}}^a_{p,s} = \left(1-(\gamma^a)^{-1} (f^a_{s,p}/f^a_{p,s})\right) $ are the non-divergent at small angles analogues of $ \mathcal{Y}^a_{p,s} = \tilde{\mathcal{Y}}^a_{p,s} \cot{\vartheta}  $ functions in Ref.~\cite{bliokh2013goos}. IF shift could be expressed as follows according to Eq.~(1):
\begin{eqnarray}
\label{Y shift}
&&
\langle Y^a \rangle
=
\frac{\langle  \tilde{\mathbf{E}}^a |i \partial / \partial k^a_Y| \tilde{\mathbf{E}}^a \rangle}
{N^a} = \frac{i}{k_0} \frac{\langle e |\,  \int_\nu \int_\mu \hat{T}^{a\dag} f(\mu,\nu) \, \partial / \partial k^a_{Y} \, f(\mu,\nu) \hat{T}^{a} d\nu d\mu  \,|e\rangle}{\langle e |\,   \int_\nu \int_\mu \hat{T}^{a\dag} f^2(\mu,\nu)   \hat{T}^{a} d\nu d\mu \,|e\rangle} 
= \frac{i}{k_0}  \dfrac{\langle e |\, \sum_{i = 1}^{i=6} I_i \,|e\rangle}{N^a} 
, 
\end{eqnarray}
where integration is taken over the transverse momentum components $k^a_X = \gamma^a k_0 \mu$ and $k^a_Y  = k_0 \nu$ in the beam coordinate frame. For the light beam incident at the air-material flat interface, we use the Gaussian field distribution with angular spectrum $f(\mu,\nu) = \exp[-\kappa^2 (\mu^2 + \nu^2)/2]$, where $\kappa=k_0 w_0/\sqrt{2}$ and $w_0$ is the beam waist. First, we focus on the matrix integral by $ \nu $ in the numerator. After matrix and Gaussian differentiation using the product rule, resulting matrix multiplication, and discarding terms linear in $ \nu $ which bring no contribution to the integral $ \int_{-\infty}^{\infty} d\nu $, we are left with six distinct terms in integration by $ \nu $ (superscripts $ ^a $ are omitted for clarity):

\begin{eqnarray}
	\label{near A}
	&& I_1 = -\kappa^2 \int_\mu d\mu \, e^{-\kappa^2 \mu^2} \, 
	\int_{-\infty}^{\infty} d\nu\, e^{-\kappa^2 \nu^2} 
	\nu\,\cos^3\phi\,\sin\phi  \cdot 
	\left(
	\begin{array}{cc}
	0  & |f^a_p|^2 \tilde{\mathcal{Y}}^a_p - |f^a_s|^2 (\tilde{\mathcal{Y}}^a_s)^{*} \\
	|f^a_p|^2 (\tilde{\mathcal{Y}}^a_p)^{*} - |f^a_s|^2 \tilde{\mathcal{Y}}^a_s & 0 \\
	\end{array}
	\right), \\
	\label{near B}
	&& I_2 = \int_\mu d\mu \,  e^{-\kappa^2 \mu^2} \, \int_{-\infty}^{\infty} d\nu\, e^{-\kappa^2 \nu^2} \
	\frac{2 \nu  \sin \phi\, \cos ^5\phi}{\theta ^2}
	\cdot 
	\left(
	\begin{array}{cc}
	0  & |f^a_s|^2 (\tilde{\mathcal{Y}}^a_s)^{*} \\
	-|f^a_p|^2 (\tilde{\mathcal{Y}}^a_p)^{*} & 0 \\
	\end{array}
	\right), \\
	\label{near C}
	&& I_3 = \int_\mu d\mu \, e^{-\kappa^2 \mu^2} \,    \int_{-\infty}^{\infty} d\nu\, e^{-\kappa^2 \nu^2} \
	\left(\frac{\cos ^4\phi}{\theta}-\frac{2 \nu ^2 \cos ^6\phi }{\theta ^3}\right)
	\cdot 
	\left(
	\begin{array}{cc}
	0  & |f^a_p|^2 \tilde{\mathcal{Y}}^a_p \\
	- |f^a_s|^2 \tilde{\mathcal{Y}}^a_s & 0 \\
	\end{array}
	\right), \\
	\label{near A_mix}
	&& I_4 = -\kappa^2 \int_\mu d\mu \, e^{-\kappa^2 \mu^2} \,
	\int_{-\infty}^{\infty} d\nu\, e^{-\kappa^2 \nu^2} 
	\nu\,\cos\phi\,\sin^3\phi \cdot 
	\left(
	\begin{array}{cc}
	0  & f^a_p f^{a*}_s \left(\tilde{\mathcal{Y}}^a_p - (\tilde{\mathcal{Y}}^a_s)^{*}\right) \\
	f^{a*}_p f^a_s \left((\tilde{\mathcal{Y}}^a_p)^{*} - \tilde{\mathcal{Y}}^a_s\right) & 0 \\
	\end{array}
	\right), \\
	\label{near C_mix1}
	&& I_5 = \int_\mu d\mu \, e^{-\kappa^2 \mu^2} \,     \int_{-\infty}^{\infty} d\nu\, e^{-\kappa^2 \nu^2} \
	\left(
	\frac{\sin ^2\phi \, \cos ^2\phi}{\theta }-\frac{2 \nu ^2 \sin ^2\phi \, \cos ^4\phi}{\theta ^3}
	\right)
	\cdot
	\left(
	\begin{array}{cc}
	0 & f^a_p f^{a*}_s \tilde{\mathcal{Y}}^a_p \\
	-f^{a*}_p f^a_s \tilde{\mathcal{Y}}^a_s & 0 \\
	\end{array}
	\right), \\
	\label{near C_mix2}
	&& I_6 = \int_\mu d\mu \, e^{-\kappa^2 \mu^2} \,    \int_{-\infty}^{\infty} d\nu\, e^{-\kappa^2 \nu^2} \
	\left(
	-\frac{2 \theta ^2 \sin ^5\phi \, \cos \phi}{\nu ^3}
	\right)
	\cdot
	\left(
	\begin{array}{cc}
	0 & f^a_p f^{a*}_s (\tilde{\mathcal{Y}}^a_s)^{*} \\
	-f^{a*}_p f^a_s (\tilde{\mathcal{Y}}^a_p)^{*} & 0 \\
	\end{array}
	\right).
\end{eqnarray}
It is worth noting that the matrices appearing in these formulas include both the standard \eqref{near A}-\eqref{near C} which already appear in approximate calculation at large incidence angles, and the emergent under near-normal incidence matrices in \eqref{near A_mix}-\eqref{near C_mix2}. 


The $ \nu-$integration using the explicit forms of trigonometric functions $ \sin\phi,\, \cos\phi $ yields the following functional multipliers attached to the corresponding matrices in \eqref{near A}-\eqref{near C_mix2}: 
\begin{eqnarray}
	\label{near A func}
	& I_1 = & \int_\mu d\mu \, e^{-\kappa^2 \mu^2} \, 
	 \left( \sqrt{\pi } \theta ^3 \kappa ^3-\frac{\pi  \theta ^2 \kappa ^2 e^{\theta ^2 \kappa ^2} \left(2 \theta ^2 \kappa ^2+1\right) (1-\text{Erf}[\kappa  \left| \theta \right| ])}{2 \, \text{sign}(\theta )} \right)
    \cdot \\  \nonumber &&
    \cdot 
	\left(
	\begin{array}{cc}
	0  & |f^a_p|^2 \tilde{\mathcal{Y}}^a_p - |f^a_s|^2 (\tilde{\mathcal{Y}}^a_s)^{*} \\
	|f^a_p|^2 (\tilde{\mathcal{Y}}^a_p)^{*} - |f^a_s|^2 \tilde{\mathcal{Y}}^a_s & 0 \\
	\end{array}
	\right), \\
	\label{near B func}
	&  I_2 = & \int_\mu d\mu \, e^{-\kappa^2 \mu^2} \, \left(
	\frac{1}{2} \sqrt{\pi } \theta  \kappa  \left(2 \theta ^2 \kappa ^2+1\right)
	-
	\frac{\pi  e^{\theta ^2 \kappa ^2} \left(4 \theta ^4 \kappa ^4+4 \theta ^2 \kappa ^2-1\right)
		(1-\text{Erf}[\kappa  \left| \theta \right| ])
	}{4 \, \text{sign}(\theta )} \right)
	\cdot \\  \nonumber &&
	\cdot 
	\left(
	\begin{array}{cc}
	0  & |f^a_s|^2 (\tilde{\mathcal{Y}}^a_s)^{*} \\
	-|f^a_p|^2 (\tilde{\mathcal{Y}}^a_p)^{*} & 0 \\
	\end{array}
	\right), \\
	\label{near C func}
	& I_3 = & \int_\mu d\mu \, e^{-\kappa^2 \mu^2} \, \left(
	\frac{1}{2} \sqrt{\pi } \theta  \kappa  \left(1-2 \theta ^2 \kappa ^2\right)
	+
	\frac{\pi  e^{\theta ^2 \kappa ^2} \left(4 \theta ^4 \kappa ^4+1\right)
		(1-\text{Erf}[\kappa  \left| \theta \right| ])
	}{4 \, \text{sign}(\theta )} \right)
	\cdot \\  \nonumber &&
	\cdot 
	\left(
	\begin{array}{cc}
	0  & f^a_p f^{a*}_s \left(\tilde{\mathcal{Y}}^a_p - (\tilde{\mathcal{Y}}^a_s)^{*}\right) \\
	f^{a*}_p f^a_s \left((\tilde{\mathcal{Y}}^a_p)^{*} - \tilde{\mathcal{Y}}^a_s\right) & 0 \\
	\end{array}
	\right), \\
	\label{near A_mix func}
	& I_4 = & - \int_\mu d\mu \, e^{-\kappa^2 \mu^2} \, \left(
	\sqrt{\pi } \theta  \kappa  \left(\theta ^2 \kappa ^2+1\right)
	+
	\frac{1}{2} \pi  \kappa ^2 \theta\left|\theta\right|  e^{\theta ^2 \kappa ^2} \left(2 \theta ^2 \kappa ^2+3\right) (1-\text{Erf}[\kappa  \left| \theta \right| ])
	\right) 
	\cdot \\  \nonumber &&
	\cdot 
	\left(
	\begin{array}{cc}
	0  & f^a_p f^{a*}_s \left(\tilde{\mathcal{Y}}^a_p - (\tilde{\mathcal{Y}}^a_s)^{*}\right) \\
	f^{a*}_p f^a_s \left((\tilde{\mathcal{Y}}^a_p)^{*} - \tilde{\mathcal{Y}}^a_s\right) & 0 \\
	\end{array}
	\right), \\
	\label{near C_mix1 func}
	& I_5 = & \int_\mu d\mu \, e^{-\kappa^2 \mu^2} \,  \left(
	\frac{1}{2} \sqrt{\pi } \theta \kappa \left(2 \theta ^2 \kappa ^2+3\right)
	-
	\frac{
		\pi  e^{\theta ^2 \kappa ^2} \left(4 \theta ^4 \kappa ^4+8 \theta ^2 \kappa ^2+1\right) 
		(1-\text{Erf}[\kappa  \left| \theta \right| ])
	}{4 \, \text{sign}(\theta )} \right)
	\cdot \\  \nonumber &&
		\cdot
	\left(
	\begin{array}{cc}
	0 & f^a_p f^{a*}_s \tilde{\mathcal{Y}}^a_p \\
	-f^{a*}_p f^a_s \tilde{\mathcal{Y}}^a_s & 0 \\
	\end{array}
	\right), \\
	\label{near C_mix2 func}
	& I_6 = & \int_\mu d\mu \, e^{-\kappa^2 \mu^2} \,  \left(
	-
	\frac{1}{2} \sqrt{\pi } \theta  \kappa  \left(2 \theta ^2 \kappa ^2+1\right)
	+
	\frac{\pi  e^{\theta ^2 \kappa ^2} \left(4 \theta ^4 \kappa ^4+4 \theta ^2 \kappa ^2-1\right)
		(1-\text{Erf}[\kappa  \left| \theta \right| ])
	}{4 \, \text{sign}(\theta )} \right)
	\cdot \\  \nonumber &&
		\cdot
	\left(
	\begin{array}{cc}
	0 & f^a_p f^{a*}_s (\tilde{\mathcal{Y}}^a_s)^{*} \\
	-f^{a*}_p f^a_s (\tilde{\mathcal{Y}}^a_p)^{*} & 0 \\
	\end{array}
	\right),
\end{eqnarray}
where $ \text{Erf}[ x ] $ is the error function of $ x $. Next, we perform the integration over $ \int_{-\infty}^{\infty} d\mu $ of \eqref{near A func}-\eqref{near C_mix2 func}, substituting $ \theta=\vartheta + \mu $. The results read
    \begin{eqnarray}
	\label{near A func full}
	&& I_1 =
	\frac{\pi  e^{-\kappa ^2 \vartheta ^2} \left(\kappa ^2 \vartheta ^2+e^{\kappa ^2 \vartheta ^2} \left(-2 \kappa ^4 \vartheta ^4+5 \kappa ^2 \vartheta ^2-6\right)+6\right)}{4 \kappa ^6 \vartheta ^5}
	\cdot 
	\left(
	\begin{array}{cc}
	0  & |f^a_p|^2 \tilde{\mathcal{Y}}^a_p - |f^a_s|^2 (\tilde{\mathcal{Y}}^a_s)^{*} \\
	|f^a_p|^2 (\tilde{\mathcal{Y}}^a_p)^{*} - |f^a_s|^2 \tilde{\mathcal{Y}}^a_s & 0 \\
	\end{array}
	\right), \\
	\label{near B func full}
	&& I_2 =
	\frac{\pi  e^{-\kappa ^2 \vartheta ^2} \left(-\kappa ^4 \vartheta ^4+2 \kappa ^2 \vartheta ^2+e^{\kappa ^2 \vartheta ^2} \left(4 \kappa ^2 \vartheta ^2-6\right)+6\right)}{4 \kappa ^6 \vartheta ^5}
	\cdot 
	\left(
	\begin{array}{cc}
	0  & |f^a_s|^2 (\tilde{\mathcal{Y}}^a_s)^{*} \\
	-|f^a_p|^2 (\tilde{\mathcal{Y}}^a_p)^{*} & 0 \\
	\end{array}
	\right), \\
	\label{near C func full}
	&& I_3 =
	\frac{\pi  e^{-\kappa ^2 \vartheta ^2} \left(-\kappa ^4 \vartheta ^4+e^{\kappa ^2 \vartheta ^2} \left(4 \kappa ^4 \vartheta ^4-6 \kappa ^2 \vartheta ^2+6\right)-6\right)}{4 \kappa ^6 \vartheta ^5}
	\cdot 
	\left(
	\begin{array}{cc}
	0  & f^a_p f^{a*}_s \left(\tilde{\mathcal{Y}}^a_p - (\tilde{\mathcal{Y}}^a_s)^{*}\right) \\
	f^{a*}_p f^a_s \left((\tilde{\mathcal{Y}}^a_p)^{*} - \tilde{\mathcal{Y}}^a_s\right) & 0 \\
	\end{array}
	\right), \\
		&& I_4 =
	-\frac{3 \pi  e^{-\kappa ^2 \vartheta ^2} \left(\kappa ^2 \vartheta ^2+e^{\kappa ^2 \vartheta ^2} \left(\kappa ^2 \vartheta ^2-2\right)+2\right)}{4 \kappa ^6 \vartheta ^5}
	\cdot 
	\left(
	\begin{array}{cc}
	0  & f^a_p f^{a*}_s \left(\tilde{\mathcal{Y}}^a_p - (\tilde{\mathcal{Y}}^a_s)^{*}\right) \\
	f^{a*}_p f^a_s \left((\tilde{\mathcal{Y}}^a_p)^{*} - \tilde{\mathcal{Y}}^a_s\right) & 0 \\
	\end{array}
	\right),
\end{eqnarray}

\begin{eqnarray}
	\label{near C_mix1 func full}
	&& I_5 =
	\frac{\pi  e^{-\kappa ^2 \vartheta ^2} \left(\kappa ^4 \vartheta ^4+4 \kappa ^2 \vartheta ^2+2 e^{\kappa ^2 \vartheta ^2} \left(\kappa ^2 \vartheta ^2-3\right)+6\right)}{4 \kappa ^6 \vartheta ^5}
		\cdot
	\left(
	\begin{array}{cc}
	0 & f^a_p f^{a*}_s \tilde{\mathcal{Y}}^a_p \\
	-f^{a*}_p f^a_s \tilde{\mathcal{Y}}^a_s & 0 \\
	\end{array}
	\right), \\
	\label{near C_mix2 func full}	
	&& I_6 =
	\frac{\pi  e^{-\kappa ^2 \vartheta ^2} \left(\kappa ^4 \vartheta ^4-2 \kappa ^2 \vartheta ^2+e^{\kappa ^2 \vartheta ^2} \left(6-4 \kappa ^2 \vartheta ^2\right)-6\right)}{4 \kappa ^6 \vartheta ^5}
		\cdot
	\left(
	\begin{array}{cc}
	0 & f^a_p f^{a*}_s (\tilde{\mathcal{Y}}^a_s)^{*} \\
	-f^{a*}_p f^a_s (\tilde{\mathcal{Y}}^a_p)^{*} & 0 \\
	\end{array}
	\right)	.
\end{eqnarray}

Finally, the nominator could be deduced as follows and decomposed into Pauli matrices $ \hat{\sigma}_1 $ and $ \hat{\sigma}_2 $, which gives a laconic result with just two contributions:
\begin{eqnarray}
	\label{Y nuremator decomposed}
	&& i \langle e |\, \sum_{i = 1}^{i=6} I_i \,|e\rangle =
		-\frac{\pi  \left(2 \kappa ^2 \vartheta ^2 - 1 +e^{-\kappa ^2 \vartheta ^2} \left(1-\kappa ^2 \vartheta ^2\right)\right)}{4 \kappa ^4 \vartheta ^3}
	\cdot  \nonumber \\
	&&  \cdot  \,
	\left(
	\langle e | \hat{\sigma}_1 | e \rangle   \cdot
	\Im[
	|f^a_p|^2 \tilde{\mathcal{Y}}^a_p
	-
	|f^a_s|^2 \tilde{\mathcal{Y}}^a_s]
	+ \langle e | \hat{\sigma}_2 | e \rangle  \cdot
	\Re[
	|f^a_p|^2 \tilde{\mathcal{Y}}^a_p
	+
	|f^a_s|^2 \tilde{\mathcal{Y}}^a_s]
	\right) \\
	&&  \nonumber
	+
	\frac{\pi  \left(1-e^{-\kappa ^2 \vartheta ^2} \left(\kappa ^2 \vartheta ^2+1\right)\right)}{4 \kappa ^4 \vartheta ^3}
	\cdot
	\left(
	\langle e | \hat{\sigma}_1 | e \rangle  \cdot
	\Im[
	f^a_p f^{a*}_s (\tilde{\mathcal{Y}}^a_p + \tilde{\mathcal{Y}}^{a*}_s)]
	+
	\langle e | \hat{\sigma}_2 | e \rangle   \cdot
	\Re[
	f^a_p f^{a*}_s (\tilde{\mathcal{Y}}^a_p + \tilde{\mathcal{Y}}^{a*}_s)]
	\right)
	.
\end{eqnarray}
By introducing the Stokes parameters of incident light as
\begin{eqnarray}
\label{Stokes 2 sigma 1}
&&
S_2 =
\langle e| \hat{\sigma}_1 | e \rangle
=
2 \Re[e_x^{*} e_y], \\
\label{Stokes 3 sigma 2}
&&
S_3 =
\langle e| \hat{\sigma}_2 | e \rangle
=
2 \Im[e_x^{*} e_y],
\end{eqnarray}
we get
\begin{eqnarray}
	\label{Y numerator RES}
	\langle  \tilde{\mathbf{E}}^a |i \frac{\partial}{\partial k^a_Y} | \tilde{\mathbf{E}}^a \rangle
	&&=
	\frac{
		\gamma^a \kappa^2 \vartheta \Lambda_1[\kappa \vartheta]}{k_0}
	\cdot
	\left(
	- S_2 \cdot
	\Im[|f^a_p|^2 \tilde{\mathcal{Y}}^a_p
	-
	|f^a_s|^2 \tilde{\mathcal{Y}}^a_s]
	-S_3 \cdot
	\Re[|f^a_p|^2 \tilde{\mathcal{Y}}^a_p
	+
	|f^a_s|^2 \tilde{\mathcal{Y}}^a_s]
	\right) +
	\nonumber \\
	&&
	+
	\frac{
		\gamma^a \kappa^2 \vartheta \Lambda_2[\kappa \vartheta]}{k_0}
	\cdot
	\left(
	S_2 \cdot
	\Im[
	f^a_p f^{a*}_s (\tilde{\mathcal{Y}}^a_p + \tilde{\mathcal{Y}}^{a*}_s)]
	+
	S_3 \cdot
	\Re[
	f^a_p f^{a*}_s (\tilde{\mathcal{Y}}^a_p + \tilde{\mathcal{Y}}^{a*}_s)]
	\right)
	,
\end{eqnarray}
where we have introduced a ``quasi-Lorentzian'' function,
\begin{eqnarray}
\label{Lambda_1}
&&
\Lambda_1[\kappa \vartheta] = 
\frac{ 2 \kappa ^2 \vartheta ^2 - 1 +e^{-\kappa ^2 \vartheta ^2} \left(1-\kappa ^2 \vartheta ^2\right)}{4 \kappa ^4 \vartheta ^4}
,
\end{eqnarray}
and a ``quasi-quad-Lorentzian'' function,
\begin{eqnarray}
\label{Lambda_2}
&&
\Lambda_2[\kappa \vartheta] = 
\frac{1-e^{-\kappa ^2 \vartheta ^2} \left(\kappa ^2 \vartheta ^2+1\right)}{4 \kappa ^4 \vartheta ^4}
.
\end{eqnarray}

The denominator (norm of the output polarization state) is calculated absolutely in the same way as nominator. Integrals arising in this calculation are those of $ \cos^4 \phi $, $ \cos^2 \phi \sin^2 \phi $, and $ \sin^4 \phi $. 
The result reads
\begin{eqnarray}
	\label{Q}
	N^a
	=
	\gamma^a
	\Big[
	&&
	F_1[\kappa \vartheta]
	\cdot
	\left(
	|f^a_p|^2|e_x|^2 + |f^a_s|^2|e_y|^2
	\right)
	+
	F_2[\kappa \vartheta]
	\cdot
	\left(
	|f^a_s|^2|e_x|^2 + |f^a_p|^2|e_y|^2
	\right)
	\nonumber\\
	&&
	+
	F_3[\kappa \vartheta]
	\cdot
	\left(
	2 \Re[f^a_p f^{a*}_s]
	+
	|f^a_s|^2 |\tilde{\mathcal{Y}}^a_s|^2 |e_x|^2 + 
	|f^a_p|^2 |\tilde{\mathcal{Y}}^a_p|^2 |e_y|^2
	\right)
	\Big]
	,
\end{eqnarray}
where we have introduced the functions
\begin{eqnarray}
\label{F1}
F_1[\kappa  \vartheta]
=
&&
\frac{e^{-\kappa ^2 \vartheta ^2} \left(\kappa ^2 \vartheta ^2-3\right)+\left(4 \kappa ^4 \vartheta ^4-4 \kappa ^2 \vartheta ^2+3\right)}{4 \kappa ^4 \vartheta ^4}
,\\
\label{F2}
F_2[\kappa \vartheta]
=
&&
\frac{3 e^{-\kappa ^2 \vartheta ^2} \left(-\kappa ^2 \vartheta ^2-1\right)+3}{4 \kappa ^4 \vartheta ^4}
,\\
\label{F3}
F_3[\kappa  \vartheta]
=
&&
\frac{e^{-\kappa ^2 \vartheta ^2} \left(\kappa ^2 \vartheta ^2+3\right)+\left(2 \kappa ^2 \vartheta ^2-3\right)}{4 \kappa ^4 \vartheta ^4}
.
\end{eqnarray}

Now we could write the result \eqref{Y shift} explicitly as a sum of two contributions, both naturally factorized into the (i) resonant geometric $G_{Y i}[\kappa \vartheta]$ and (ii) material-specific components $\mathcal{F}^a_{Y i}[f^a_{s/p}, S_2, S_3]$:
\begin{eqnarray}
\label{Y RES short}
\langle Y^a \rangle \cdot k
&&= 
\sum_{i=1}^{2}
G_{Y i}[\kappa \vartheta] \cdot 
\mathcal{F}^a_{Y i}[f^a_{s/p}, S_2, S_3]
,
\end{eqnarray}
where
\begin{eqnarray}
\label{G}
&& G_{Y 1}[\kappa \vartheta] =
\kappa^2 \vartheta \cdot \Lambda_1[\kappa \vartheta], \; G_{Y 2}[\kappa \vartheta] =
\kappa^2 \vartheta \cdot \Lambda_2[\kappa \vartheta]  \\
\label{F_Y1}
&&
\mathcal{F}^a_{Y 1}
=
\frac{
	-S_2 \cdot
	\Im[|f^a_p|^2 \tilde{\mathcal{Y}}^a_p
	-
	|f^a_s|^2 \tilde{\mathcal{Y}}^a_s]
	-S_3 \cdot
	\Re[|f^a_p|^2 \tilde{\mathcal{Y}}^a_p
	+
	|f^a_s|^2 \tilde{\mathcal{Y}}^a_s]
}
{
	N^a
}
,  \\
\label{F_Y2}
&&
\mathcal{F}^a_{Y 2}
=
\frac
{
	S_2 \cdot
	\Im[
	f^a_p f^{a*}_s (\tilde{\mathcal{Y}}^a_p + \tilde{\mathcal{Y}}^{a*}_s)]
	+
	S_3 \cdot
	\Re[
	f^a_p f^{a*}_s (\tilde{\mathcal{Y}}^a_p + \tilde{\mathcal{Y}}^{a*}_s)]
}
{
N^a
}.
\end{eqnarray}

In isotropic materials, the Fresnel amplitudes $ f^a_p $, $ f^a_s $ split quadratically at small $ \vartheta $, $ \tilde{\mathcal{Y}}^a_{s,p}\propto \vartheta^2 $. Thus, we could set $ \tilde{\mathcal{Y}}^a_{s,p} \to 0 $, $ f^a_s \approx f^a_p $, which gives, in the leading-order approximation, 
\begin{eqnarray}
\label{Qsim}
&& N^a
\approx \,
\gamma^a
\cdot
\left(
|f^a_p|^2|e_x|^2 + |f^a_s|^2|e_y|^2
\right)
, \\
\label{F_Y1_simpl}
&&
\mathcal{F}^{a\,\text{isotr}}_{Y 1}
\approx
-S_2 \cdot
\Im[\tilde{\mathcal{Y}}^a_p
-
\tilde{\mathcal{Y}}^a_s]
-S_3 \cdot
\Re[\tilde{\mathcal{Y}}^a_p
+
\tilde{\mathcal{Y}}^a_s]
, \\
\label{F_Y2_simpl}
&&
\mathcal{F}^{a\,\text{isotr}}_{Y 2}
\approx
S_2 \cdot
\Im[
\tilde{\mathcal{Y}}^a_p + \tilde{\mathcal{Y}}^{a*}_s]
+
S_3 \cdot
\Re[
\tilde{\mathcal{Y}}^a_p + \tilde{\mathcal{Y}}^{a*}_s]
.
\end{eqnarray}
Note that Eq.~\eqref{Qsim} is perfectly consistent with expression for the proper normalization in Ref.~\cite{bliokh2013goos} at large $ \vartheta $.

\begin{figure}[h!]
	\includegraphics [width=1.0\textwidth]{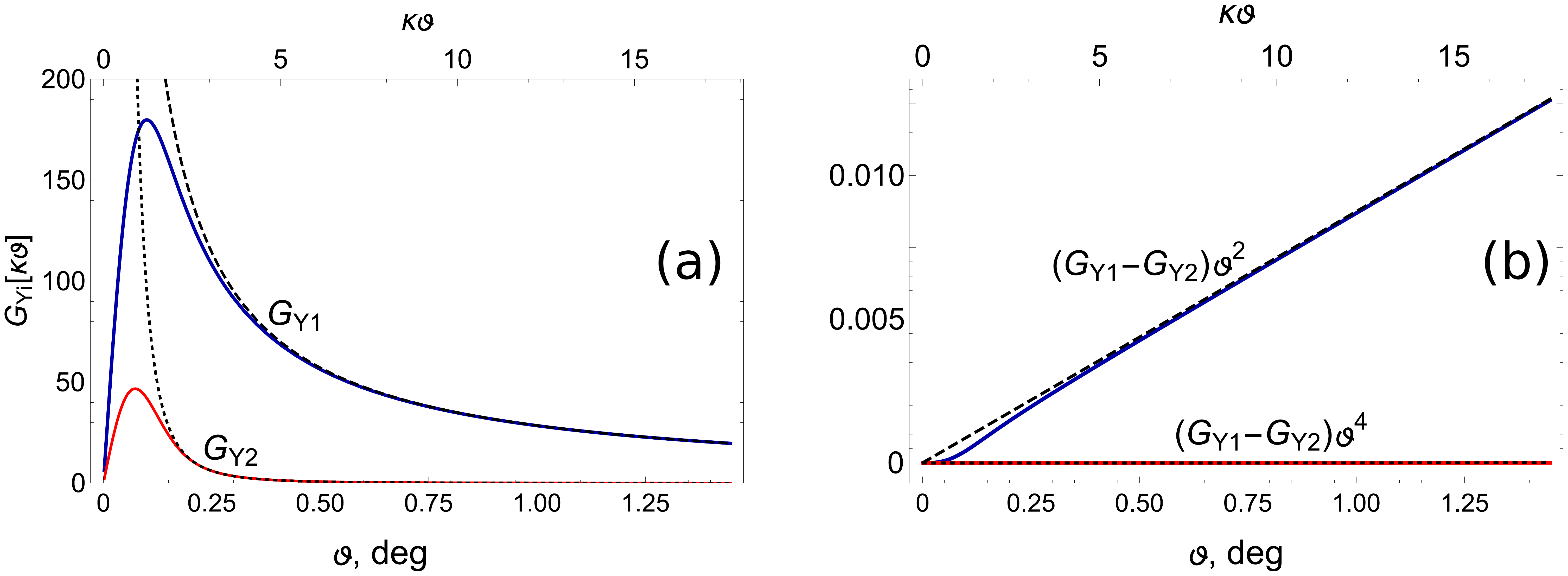}
	\caption{
		\textit{Panel (a):} the resonant geometric factors \eqref{G} $ G_{Y 1}[\kappa \vartheta] $ (solid blue) and $ G_{Y 2}[\kappa \vartheta] $ (solid red), and the asymptotes $ 1 /(2 \vartheta ) $ (black dashed) and $1/(4\kappa^2 \vartheta^3)$ (black dotted) at large angles $\kappa \vartheta\gg 1 $. 
		\textit{Panel (b):} the non-resonant angular dependence of near-normal shifts $ (G_{Y 1} - G_{Y 2})\cdot \vartheta^2 $ (solid blue) and $ (G_{Y 1} - G_{Y 2})\cdot \vartheta^4 $ (solid red), and the corresponding asymptotes $ \vartheta/2 $ (black dashed) and $ \vartheta^3/2 $ (black dotted) at large angles $\kappa \vartheta\gg 1 $. 
		Here, we use $\kappa = 700$.
	}\label{Fig G}
\end{figure}

The material-dependent terms $\mathcal{F}^{a\,\text{isotr}}_{Y i}$ are proportional to $(\vartheta^2 + \mathcal{O}(\vartheta^4))$, while the geometric resonant terms  $ G_{Y i}[\kappa \vartheta] $ exhibit the sharp peak under $\kappa \vartheta \sim 1$. The asymptotics of the geometric resonant terms $G_{Y 1}[\kappa \vartheta]$ and $G_{Y 2}[\kappa \vartheta]$ at the large angles corresponding to $\kappa \vartheta \gg 1$ are $1/(2\vartheta)$ and $1/(4\kappa^2 \vartheta^3)$, respectively. So, the asymptote of $G_{Y 1}[\kappa \vartheta]$ is dominant at large angles and is consistent with the standard theory~\cite{bliokh2013goos}. The proper resonant dependences of the geometric factors \eqref{G} alongside with various asymptotics are depicted in Fig.~\ref{Fig G}. 

Next we examine the material-dependent components. For isotropic materials, the Fresnel coefficients $ f^a_{s,p} $ are even and quadratic in small $ \vartheta $. Thus, we pick a generic model $ f^a_{s,p}=f_0 + \alpha_{s,p} \vartheta^2 + \beta_{s,p} \vartheta^4 $, from which follows
\begin{eqnarray}
\label{Ys isotropic}
&&
\tilde{\mathcal{Y}}_p
-
\tilde{\mathcal{Y}}_s
=
\frac{2(\alpha_p-\alpha_s)}{f_0} \vartheta^2 + \mathcal{O}(\vartheta^4)
, \\
&&
\tilde{\mathcal{Y}}_p
+
\tilde{\mathcal{Y}}_s
=
-\frac{(\alpha_p-\alpha_s)^2}{f^2_0} \vartheta^4 + \mathcal{O}(\vartheta^6)
, \\
&&
\tilde{\mathcal{Y}}_p + \tilde{\mathcal{Y}}^{*}_s
=
i
\Im\left[\frac{2 (\alpha_p-\alpha_s)}{f_0}\right] \vartheta^2
+
\mathcal{O}(\vartheta^4)
.
\end{eqnarray}
Thus, the $ S_2$-parts of both \eqref{F_Y1_simpl} and \eqref{F_Y2_simpl} are $ \propto \Im\left[2 (\alpha_p-\alpha_s)/f_0\right]\vartheta^2 $, thus $ \langle Y \rangle \propto(G_{Y 1}-G_{Y 2})\vartheta^2 $, which is almost linear dependence, and thus there is no peak (see Fig.~\ref{Fig G}, panel (b), blue curve). 
The $ S_3$-parts of both \eqref{F_Y1_simpl} and \eqref{F_Y2_simpl} are $ \propto \Re\left[(\alpha_p-\alpha_s)^2/f_0^2\right]\vartheta^4 $, thus $ \langle Y \rangle \propto(G_{Y 1}-G_{Y 2})\vartheta^4 $, and the shift is again not peaked, being cubic in small angles (see Fig.~\ref{Fig G}, panel (b), red curve).

The analogous procedure could be conducted for another types of shifts, namely $\langle X \rangle$, $\langle P_X \rangle$ and $\langle P_Y \rangle$, leading to the similar suppression of the geometric resonance by the material-dependent term proportional to $[\vartheta^2 +\mathcal{O}(\vartheta^2)]$ at near-normal incidence. Thus, the anomalous optical beam shifts cannot be observed for isotropic medium and require the finite difference between the Fresnel amplitudes at strictly normal incidence.


\section{Supplemental Note 5: Near-normal shifts in uniaxial waveplate}

In this section, similarly to the previous one, we briefly go through the calculations of all the shifts for the anisotropic waveplate. The Jones matrix for the waveplate has already been discussed in detail in Supplemental Note~3. 
As in the main text, we calculate the shifts for the transmitted beam, while the calculations for reflected beam could be done analogously. 

We first calculate the normalization $N^t$ which appears in the denominators of all of the shifts. The matrix product appearing under the integral in the calculation of the normalization reads 
\begin{eqnarray}
	\label{TT_uniax}
	\hat{T}^{t \, \dag}	\hat{T}^t = \frac{1}{2}	\left(
	\begin{array}{cc}
	\tau_+ + \tau_- \cos (4 \phi ) & \tau_- \sin (4 \phi ) \\
	\tau_- \sin (4 \phi ) & \tau_+ - \tau_- \cos (4 \phi ) \\
	\end{array}
	\right)
	= \frac{1}{2} \left(
	\tau_+ \hat{\sigma}_1 + \tau_- \cos (4 \phi ) \hat{\sigma}_3 + \tau_- \sin (4 \phi ) \hat{\sigma}_2 \right)
	. 
\end{eqnarray}
After the integration over $\mu$ and $\nu$ according to Eq.~(1) we get 
\begin{equation}
\label{xi s3}
     N^t = \frac{1}{2} \tau_+ \langle e | \hat{\sigma}_1 | e \rangle + \tau_- \langle e | \hat{\sigma}_3 | e \rangle \cdot 
     \frac{6 - 4 \kappa^2 \vartheta^2 + \kappa^4 \vartheta^4 -2 e^{-\kappa^2 \vartheta^2} \left(\kappa^2 \vartheta^2 + 3\right)}{2\kappa^4 \vartheta^4}
    .
\end{equation}
The latter result is obtained after noting that $\cos(4 \phi) = \cos^4\phi - 6\sin^2\phi \cos^2\phi + \sin^4\phi$, which after integration gives the corresponding terms 
$e^{- \kappa ^2 \vartheta ^2} \left(\kappa ^2 \vartheta ^2+e^{\kappa ^2 \vartheta ^2} \left(4 \kappa ^4 \vartheta ^4-4 \kappa ^2 \vartheta ^2+3\right)-3\right) \Big/ (4 \kappa ^4 \vartheta ^4)$, 
$e^{- \kappa ^2 \vartheta ^2} \left(\kappa ^2 \vartheta ^2+e^{\kappa ^2 \vartheta ^2} \left(2 \kappa ^2 \vartheta ^2-3\right)+3\right) \Big/ (4 \kappa ^4 \vartheta ^4)$, 
and $3 e^{- \kappa ^2 \vartheta ^2} \left(- \kappa ^2 \vartheta ^2 + e^{\kappa ^2 \vartheta ^2}-1\right) \Big/ (4 \kappa ^4 \vartheta ^4)$. 
Calculating the polarization expectation values in Eq.~\eqref{xi s3}, we come to the result from the main text: $ 2 N^t = \tau_+ + \tau_- \, S_1 \, \Lambda_Y(\kappa \vartheta)$, where $\Lambda_Y(\kappa \vartheta)$ is the nonlinear function appearing in~\eqref{xi s3} (also mentioned in the main text). 

In analogous manner, the nominators for all the shifts are calculated. For completeness, we mention below the corresponding integrands and term-by-term results of integration, as in the case of the normalization $ N^t $. 

\

Angular GH shift: $\langle P_X \rangle \cdot \kappa^2/k_0$. 
The matrix product under the integral: $\mu \cdot \hat{T}^{t \, \dag} \hat{T}^t $. 
By parity, only diagonal terms $\tau_- \mu \cos (4 \phi ) \hat{\sigma}_3 / 2$ survive. 
After integration, the terms coming from the expansion of $\cos(4 \phi) = \cos^4\phi - 6\sin^2\phi \cos^2\phi + \sin^4\phi$ give, respectively, 
$ e^{- \kappa ^2 \vartheta ^2} \left(- \kappa ^4 \vartheta ^4 + 2 \kappa ^2 \vartheta ^2+e^{\kappa ^2 \vartheta ^2} \left(4 \kappa ^2 \vartheta ^2-6\right)+6\right) \Big/ (4 \kappa ^6 \vartheta ^5)$, 
$\left(-2 \kappa ^2 \vartheta ^2-e^{- \kappa ^2 \vartheta ^2} \left(\kappa ^4 \vartheta ^4+4 \kappa ^2 \vartheta ^2+6\right)+6\right) \Big/ (4 \kappa ^6 \vartheta ^5)$, 
and $3 e^{- \kappa ^2 \vartheta ^2} \left(\kappa ^4 \vartheta ^4+2 \kappa ^2 \vartheta ^2-2 e^{\kappa ^2 \vartheta ^2}+2\right) \Big/ (4 \kappa ^6 \vartheta ^5)$. 
Summing up the results and surrounding them by the bra- and ket-states of incident light polarization, we get Eq.~(4) from the main text. 

\

Angular IF shift: $\langle P_Y \rangle \cdot \kappa^2/k_0$. 
The matrix product under the integral: $\nu \cdot \hat{T}^{t \, \dag} \hat{T}^t $. By parity, only off-diagonal terms $\tau_- \nu \sin (4 \phi ) \hat{\sigma}_1 /2$ survive. In integration, we use $\sin (4 \phi ) = 4 \left(\sin\phi \cos^3\phi - \sin^3\phi \cos\phi\right)$. After integration, we get 
$4 \tau_- \hat{\sigma}_1 \left(\kappa ^4 \vartheta ^4-4 \kappa ^2 \vartheta ^2-2 e^{\kappa ^2 \left(-\vartheta ^2\right)} \left(\kappa ^2 \vartheta ^2+3\right)+6\right) \Big/ (\kappa ^6 \vartheta ^5)$, which finally leads to Eq.~(5). 

\

Linear GH shift: $\langle X \rangle \cdot k_0$. 
The matrix product under the integral: $ i \hat{T}^{t \, \dag} (\partial \hat{T}^t / \partial \mu) - i \kappa^2 \mu \cdot \hat{T}^{t \, \dag} \hat{T}^t $. 
The second part of this expression (taken apart from $- \kappa^2$) already appeared in the calculation of $\langle P_X \rangle$ and thus leads to the proportional term, $- i S_1 \cdot \tau_- \cdot  \Lambda_X(\kappa \vartheta) / \vartheta $. 
The first part leads to the non-vanishing by parity integrand $ i \hat{\sigma}_3 \left( \left| t_- \right|^2 + t_- t_{+}^* \right) \nu \cos^2(\phi ) \sin(4 \phi) / (\vartheta + \mu)^2 $, which after integration gives $i S_1 \left( \left| t_- \right|^2 + t_- t_{+}^* \right) \Lambda_X(\kappa \vartheta) / \vartheta $. After adding these contributions, again noting that $t_- t_{+}^* = \left( \tau_- + i \tau_\times \right)$, and discarding imaginary terms, we obtain Eq.~(6) from the main text. 

\

Linear IF shift: $\langle Y \rangle \cdot k_0$ . 
The matrix product under the integral: $ i \hat{T}^{t \, \dag} (\partial \hat{T}^t / \partial \nu) - i \kappa^2 \nu \cdot \hat{T}^{t \, \dag} \hat{T}^t $. 
The second term of this expression (taken apart from $- i \kappa^2$) has already appeared in the calculation of angular IF shift $\langle P_Y \rangle$, and thus leads to the proportional term, $- i S_2 \cdot \tau_- \cdot  \Lambda_Y(\kappa \vartheta) / \vartheta $. 
The first part leads to the non-vanishing by parity integrand $ - \hat{\sigma}_2 \left| t_- \right|^2 \cos^2(\phi ) /  (\vartheta + \mu)  +  \hat{\sigma}_1 \left( i \left| t_- \right|^2 + i t_- t_{+}^* \right) \cos^2(\phi ) \cos (4 \phi ) / (\vartheta + \mu) $, which after integration gives $ - S_3 \left| t_- \right|^2 \left(1-e^{ - \kappa^2 \vartheta^2}\right)/ \vartheta + i S_2 \left(  \left| t_- \right|^2 + t_- t_{+}^* \right) \Lambda_Y(\kappa \vartheta) / \vartheta $. After adding these contributions, noting that $t_- t_{+}^* = \left( \tau_- + i \tau_\times \right) $, and discarding imaginary terms, we obtain Eq.~(7) from the main text.

\bibliography{refs}